\pdfoutput=1
\documentclass[11pt]{article}
\usepackage{jheppub}
\usepackage{amsmath, amsthm}

\usepackage{tikz}
\usepackage{tikz-cd}
\usepackage{slashed}
\usetikzlibrary{matrix,arrows}
\usepackage{multirow}

\newcommand{\eq}[1]{(\ref{#1})}
\usepackage{pifont}

\begin{document}

\title{Bounds on Discrete Gauge Symmetries in Supergravity}

\author[1]{Zihni Kaan Baykara,}
\author[2]{Markus Dierigl,}
\author[3]{Hee-Cheol Kim,}
\author[1]{Cumrun Vafa,}
\author[1]{Kai Xu}
\affiliation[1]{Jefferson Physical Laboratory, Harvard University,\\
Cambridge, MA 02138, USA}
\affiliation[2]{Theoretical Physics Department, CERN, 1211 Geneva 23, Switzerland}
\affiliation[3]{Department of Physics, POSTECH, Pohang 37673, Korea}

\preprint{CERN-TH-2025-233}

\abstract{We place bounds on the order of enhanced discrete gauge symmetries that act on massless fields and thus arise at subloci of the moduli space in supergravity theories.  We focus on supersymmetric theories with 8 or more supercharges which in some cases lead to sharp upper bounds realized by specific string constructions.}
\maketitle

\section{Introduction}

In recent years we have learned that coupling a gauge theory to gravity imposes many consistency conditions on the realization of gauge groups. These range from a bound on the lightest charged particle \cite{Arkani-Hamed:2006emk} to bounds on the rank of non-Abelian gauge factors in supergravity \cite{Kim:2019ths}, and have been formulated in terms of a set of Swampland Conjectures \cite{Vafa:2005ui}.\footnote{For other related work see, e.g., \cite{Banks:2010zn,Harlow:2015lma,Heidenreich:2016aqi,Harlow:2018jwu,Lee:2019skh,Katz:2020ewz,Hamada:2021bbz,Martucci:2022krl}, and \cite{Palti:2019pca, vanBeest:2021lhn, Agmon:2022thq} for reviews.} Most of these bounds, however, focus on continuous gauge symmetries. Discrete symmetries have not been addressed, even though from the Swampland Program's perspective we expect them to be bounded. Indeed if this were not the case, it would give another infinity to get rid of in the quantum gravity landscape, e.g., \cite{Kumar:2010ru, Hamada:2021yxy, Grimm:2023lrf, Kim:2024eoa, Delgado:2024skw, Hamada:2025vga, Birkar:2025rcg}.  Even if we prove it is bounded, finding a good bound would be rather useful as this often enters various other Swampland conjectures.  For example the index of the sublattice in variants of the weak gravity conjecture require such bounds \cite{Heidenreich:2015nta, Heidenreich:2016aqi}. In this work we therefore aim to bound discrete gauge symmetries in supergravity theories with 32, 16, and 8 supercharges.

We mainly focus on discrete gauge symmetries that act on the massless moduli fields of the supergravity theory.  In other words, we are interested in finding bounds on discrete gauge symmetries that arise at subloci of moduli space and can be Higgsed as we move away from such loci by giving vacuum expectation value to charged massless fields.  Moreover, in favorable cases, through the dependence of the properties of BPS states on the moduli, they also act on the integral lattice of BPS states and therefore need to satisfy charge quantization conditions. 

We are particularly interested in the realization of cyclic gauge groups $\mathbb{Z}_N$ with the maximal order $N$ (as well as maximal prime order $p$), as a measure of the maximal discrete gauge symmetries that are allowed. To do so, we use our knowledge of duality symmetries in various supergravity theories in $D \leq 9$ dimensions. In particular, we use their duality groups. Taking into account the quantization of charges, these are given by:
\begin{itemize}
    \item{32 supercharges: U-duality groups $G^D_{\mathrm{U}}(\mathbb{Z})$}
    \item{16 supercharges: T-duality groups SO$(26-D,10-D; \mathbb{Z})$}
    \item{8 supercharges: Duality in tensor multiplet sector SO$(1,T;\mathbb{Z})$ }
\end{itemize}
For theories with 16 supercharges, even though there are other duality groups for reduced rank cases, we show why the maximal rank given above also leads to maximal discrete gauge symmetry.
For 8 supercharges we improve the existing supergravity bounds on the rank of the gauge group and the number of neutral hypermultiplets. In these theories there can also be enhanced gauge symmetries on hypermultiplet moduli space.  However, the hypers do not control masses of charged BPS states and so our techniques are inadequate for that case. Nevertheless, we make some progress with it assuming permutation type symmetries and also leverage anomalies to restrict it. For the rest
we will bound or derive the maximal cyclic subgroups of these groups. We further discuss the specific points in moduli space, where large discrete symmetries are unbroken. Our bounds for theories with 16 and 32 supercharges are summarized in Table \ref{tab:results}.\footnote{Note that it is surprising that the maximal prime order bound for $N=16$ is stricter than what can be realized with 32 supercharges in $D=3$ dimensions. Here, we want to stress that the $N=16$ bound refers to theories with exactly 16 supercharges and thus excludes the $N =32$ case. Furthermore, for $N=32$ in $D=3$ there is no lattice of BPS states which might make the interpretation of the bounds more subtle.}
\begin{table}
\centering
    \begin{tabular}{| c || c | c |}
    \hline
    $D$ & $N = 32$ & $N = 16$ \\ \hline \hline
    $9$ & $\mathbb{Z}_p = \mathbb{Z}_3 \,, \enspace \mathbb{Z}_N = \mathbb{Z}_6$ & $\mathbb{Z}_p = \mathbb{Z}_{17} \,, \enspace \mathbb{Z}_N = \mathbb{Z}_{840} $ \\ \hline
    $8$ & $\mathbb{Z}_p = \mathbb{Z}_3\,, \enspace \mathbb{Z}_N = \mathbb{Z}_{12}$ & $\mathbb{Z}_p = \mathbb{Z}_{19} \,, \enspace \mathbb{Z}_N = \mathbb{Z}_{2520}$ \\ \hline
    $7$ & $\mathbb{Z}_p = \mathbb{Z}_5 \,, \enspace \mathbb{Z}_N = \mathbb{Z}_{12} $ & $\mathbb{Z}_p = \mathbb{Z}_{19} \,, \enspace \mathbb{Z}_N = \mathbb{Z}_{2520}$ \\ \hline
    $6$ & $\mathbb{Z}_p = \mathbb{Z}_7 \,, \enspace \mathbb{Z}_N = \mathbb{Z}_{60} $ & $\mathbb{Z}_p = \mathbb{Z}_{23} \,, \enspace \mathbb{Z}_N = \mathbb{Z}_{5040}$  \\ 
    \hline
    $5$ & $\mathbb{Z}_p = \mathbb{Z}_{13} \,, \enspace \mathbb{Z}_N = \mathbb{Z}_{60} $ & $\mathbb{Z}_p = \mathbb{Z}_{23} \,, \enspace \mathbb{Z}_N = \mathbb{Z}_{5040}$ \\ \hline
    $4$ & $\mathbb{Z}_p = \mathbb{Z}_{19} \,, \enspace \mathbb{Z}_N = \mathbb{Z}_{240} $ & $\mathbb{Z}_p = \mathbb{Z}_{23} \,, \enspace \mathbb{Z}_N = \mathbb{Z}_{13860}$ \\ \hline
    $3$ & $\mathbb{Z}_p = \mathbb{Z}_{31} \,, \enspace \mathbb{Z}_N = \mathbb{Z}_{6930} $ & $\mathbb{Z}_p = \mathbb{Z}_{23} \,, \enspace \mathbb{Z}_N = \mathbb{Z}_{13860}$ \\ \hline
    \end{tabular}
    \caption{Summary of the bounds on maximal discrete gauge groups in supergravity theories with 32 and 16 supercharges. All prime order bounds (except potentially in $D=5$) for 32 supercharges and maximal order bounds for $D \geq 5$ can be realized in string constructions. For the remaining cases the bounds may not be sharp and in the case of maximal order $N$ for 32 supercharges in dimensions $D\leq4$ depend on an assumption explained in Appendix~\ref{app:compact}.}
    \label{tab:results}
\end{table}

The manuscript is organized as follows. In Section \ref{sec:strategy} we outline our general approach and elucidate it in a very simple example. In Sections \ref{sec:32}, \ref{sec:16}, \ref{sec:8} we derive the bounds for discrete gauge symmetries for supergravity theories with 32, 16, and 8 supercharges, respectively. For the 8 supercharge case, while we mainly focus on the discrete symmetries in the tensor multiplet sector in Section \ref{sec:8}, we also describe restrictions of discrete symmetries acting on the hypermultiplets via permutation in Section \ref{sec:hyper} and charged hypers in Section \ref{subsec:discanom}. We conclude and point out future directions for research to generalize our results in Section \ref{sec:concl}. Some details for the computations and a discussion of a relation to elements in the maximal compact subgroup of U-dualities, are presented in Appendix \ref{app:compact}; the six-dimensional anomaly cancellation conditions for supergravity theories with 8 supercharges are summarized in Appendix \ref{app:6danomaly}.

\section{Bounding discrete symmetries}
\label{sec:strategy}

In this section we briefly outline our general strategy to bound discrete symmetries in supergravity with various amounts of supersymmetry. In particular, we are interested in discrete symmetries that act on the scalar fields (moduli) of the supergravity theory. Since these scalar fields control the value of the central charges of BPS states, and therefore their mass or tension, the symmetries must also act on the BPS states themselves. At generic points of the moduli space the symmetries will be broken spontaneously, but subgroups can be restored at special loci. At these loci the action leaves the vacuum expectation value of the moduli fields invariant, but continues to act on the BPS states. However, this means that the discrete symmetries need to satisfy the associated quantization condition for the BPS charges and act as general linear transformations over the integers, i.e., elements of GL$(n;\mathbb{Z})$. From this fact alone we will see that we get a bound on the maximal (prime) order of elements of the discrete groups. We further want to focus on discrete symmetries that are not part of continuous symmetries. Every continuous gauge symmetry has arbitrary $\mathbb{Z}_N$ subgroups. Similarly Weyl groups of non-Abelian gauge factors can be excluded.

\subsection{A rough upper bound}
\label{subsec:rough}

Given this realization of discrete symmetries as subgroups of the general linear group over the integers, we can already obtain universal upper bounds. Consider a symmetry acting on the moduli which is only preserved on a subspace of the corresponding moduli space. Moreover, let us assume that the central charges of some charged BPS branes depend on the moduli. In our examples the branes will be BPS but the argument we are presenting here is more general and only depends on the assumption of the existence of integral charged branes whose mass/tension varies with the moduli fields. Let the rank of the corresponding charge lattice be given by $n$. Then any symmetry which is preserved only on some proper subspace of this moduli space must act on the corresponding lattice; if it were not to act on the charge lattice that would imply that as we go infinitesimally away from this symmetry locus on the moduli space the BPS masses do not change, in contradiction with our assumption. Since the charged states are integrally quantized this implies that the symmetry group should be a subgroup of GL$(n;\mathbb{Z})$. Given this fact, we can use cyclotomic polynomials and their companion matrices to obtain order $N$ elements in GL$(n;\mathbb{Z})$ to place bounds on the order of symmetry group elements. Let us recall some elementary definitions, for more detailed accounts see \cite{Baykara:2024vss, Harvey:1987da, newman, 5e076f03-1178-3b70-95a5-ab41b5a4d470}.

The cyclotomic polynomials are defined as
\begin{equation}
    \Phi_m (x) = \prod_{\substack{0 < r < m \\ \mathrm{gcd}(r,m) = 1}} (x - \xi^{r}) = x^{\phi(m)} + a_{\phi(m) - 1} x^{\phi(m) - 1} + \dots + a_1 x + a_0 \,, \quad \xi = e^{2 \pi i / m} \,,
    \label{eq:cyclo}
\end{equation}
with Euler totient $\phi(m)$ counting integers in $\{1 \,, 2 \,, \dots, m-1 \}$ that are coprime to $m$. The integer coefficients $a_i$ define the associated companion matrix, which is given by
\begin{equation}
    C (\Phi_m ) = \begin{pmatrix} 0 & 0 & 0 & \dots & 0 & -a_0 \\ 1 & 0 & 0&  \dots & 0 & - a_1 \\ 0 & 1 & 0 & \dots & 0 & -a_2 \\ \vdots & & & & & \vdots \\ 0 & 0 & 0 & \dots & 1 & -a_{\phi(m)-1}\end{pmatrix} \in \mathrm{GL}\big(\phi(m); \mathbb{Z} \big) \,.
\end{equation}
The eigenvalues of this matrix are precisely given by the $\xi^r$ in \eqref{eq:cyclo}. By construction this matrix forms an element of order $N$ and is irreducible (over the integers). Thus given a fixed rank $n$, we can bound the maximal finite order elements by combining these companion matrices
\begin{equation}
    \begin{pmatrix} C(\Phi_{m_1}) & 0 & \dots & 0 \\ 0 & C (\Phi_{m_2}) & & 0 \\ 0 & 0 & \dots & C(\Phi_{m_r})\end{pmatrix} \in \mathrm{GL}(n;\mathbb{Z}) \,,
    \label{eq:blocks}
\end{equation}
which forms an element of order
\begin{equation}
   N = \mathrm{lcm}(m_1 \,, m_2 \,, \dots \,, m_r) \,.
\end{equation}
With this construction we can find elements of order $N$ with prime decomposition 
\begin{equation}
    N = p_1^{s_1} \, p_2^{s_2} \, \dots p_{\ell}^{s_{\ell}} \,,
\end{equation}
in GL$(n;\mathbb{Z})$ if and only if, \cite{5e076f03-1178-3b70-95a5-ab41b5a4d470},
\begin{equation}
\begin{split}
    &\sum_{j = 1}^{\ell} (p_j -1)p_j^{s_j-1} -1 \leq n \,, \quad p_1^{s_1} = 2 \,, \\
    & \sum_{j=1}^{\ell} (p_j -1)p_j^{s_j-1} \leq n \,, \quad \text{otherwise} \,.
\end{split}
\end{equation}
Consequently, in order to find the largest order of a cyclic subgroup, we have to maximize $N$ for fixed $n$.

For $N = p$ prime (and let us assume $p>2$) we see that the formula above reads
\begin{equation}
    p-1 \leq n \,,
\end{equation}
and thus all primes smaller than or equal to $n+1$ can be produced in GL$(n;\mathbb{Z})$. 

For maximizing $N$ we can optimize the sizes of the individual building blocks as in \eqref{eq:blocks}, to get the largest order possible. This leads to 
\begin{equation}
    \begin{array}{c || c | c | c | c | c | c | c | c | c | c | c | c | c | c | }
n & 2 & 4 & 6 & 8 & 10 & 12 & 14 & 16 & 18 & 20 & 22 & 24 & 26 & 28 \\ \hline
\mathbb{Z}_N & \mathbb{Z}_6 & \mathbb{Z}_{12} & \mathbb{Z}_{30} & \mathbb{Z}_{60} & \mathbb{Z}_{120} & \mathbb{Z}_{210} & \mathbb{Z}_{420} & \mathbb{Z}_{840} & \mathbb{Z}_{1260} & \mathbb{Z}_{2520} & \mathbb{Z}_{2520} & \mathbb{Z}_{5040} & \mathbb{Z}_{9240} & \mathbb{Z}_{13860}
\end{array}
\label{eq:GLmax}
\end{equation}
for larger $n$, see \cite{LEVITT1998630}.

While this procedure leads to rough upper bounds, we can refine these using additional constraints in setups with a large number of supercharges. These can sharpen the upper bounds significantly with respect to the maximal elements in GL$(n,\mathbb{Z})$ above and will be explored in Section \ref{sec:32}, \ref{sec:16}, and \ref{sec:8}, where we also argue whether we expect the upper bounds to be realized in string theory.

\subsection{A simple example}

\begin{figure}
    \centering
    \includegraphics[width=0.6\linewidth]{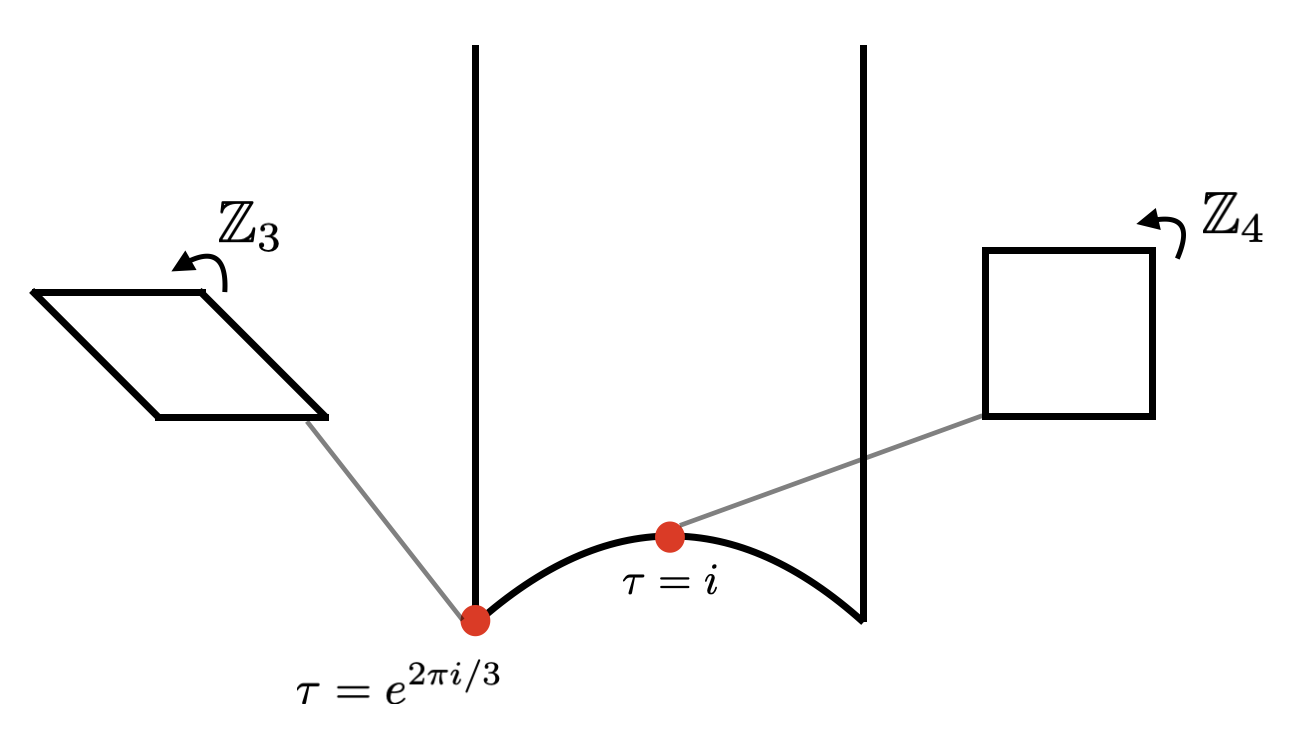}
    \caption{Points in moduli space (of $\tau$) at which a $\mathbb{Z}_3$ and $\mathbb{Z}_4$ symmetry of the full SL$(2;\mathbb{Z})$ duality are restored.}
    \label{fig:enhancementpoints}
\end{figure}

Let us discuss this phenomenon in a simple example; maximally supersymmetric ($32$ real supercharges) supergravity theory in nine dimensions. This theory has a U-duality group that contains SL$(2;\mathbb{Z})$. In its realization as M-theory on a 2-torus this SL$(2;\mathbb{Z})$ is associated with large diffeomorphisms. The nine-dimensional supergravity theory has BPS strings that come from M2-branes wrapping 1-cycles of the compactification torus (we choose the usual A- and B-cycles as a basis for $H_1(T^2;\mathbb{Z})$). At fixed volume, the torus is described by a single complex scalar field $\tau$, the complex structure. The vacuum expectation value of this complex scalar field generically breaks the SL$(2;\mathbb{Z})$ symmetry spontaneously, under which 
\begin{equation}
    \tau \mapsto \frac{a \tau  + b}{c \tau +d} \,, \quad \begin{pmatrix} a & b \\ c & d \end{pmatrix} \in \text{SL}(2;\mathbb{Z}) \,.
\end{equation}
The BPS strings also transform under these transformations\footnote{Here, we keep the product $\Sigma_B C_B + \Sigma_A C_A$ between strings and 2-forms invariant, motivated by the duality to F-theory, see e.g., \cite{Weigand:2018rez}.}
\begin{equation}
    \begin{pmatrix} \Sigma_B \\ \Sigma_A \end{pmatrix}^T \mapsto \begin{pmatrix} \Sigma_B \\ \Sigma_A \end{pmatrix}^T \begin{pmatrix} d & -b \\ -c & a \end{pmatrix} \,.
\label{eq:stringtrafo}
\end{equation}
However, at particular values of $\tau$, a finite subgroup of SL$(2;\mathbb{Z})$ can be restored, e.g.,
\begin{equation}
\begin{split}
\tau = i:& \quad \mathbb{Z}_4 \subset \mathrm{SL}(2;\mathbb{Z}) \,, \quad \begin{pmatrix} 0 & -1 \\ 1 & 0 \end{pmatrix} \,, \\
\tau = e^{2 \pi i/3}:& \quad \mathbb{Z}_3 \subset \mathrm{SL}(2;\mathbb{Z}) \,, \quad \begin{pmatrix} -1 & -1 \\ 1 & 0 \end{pmatrix} \,.
\end{split}
\label{eq:SL2matrices}
\end{equation}
At these values the string tensions of generators in a suitable basis coincide and they are acted upon by the $\mathbb{Z}_k$ transformations according to \eqref{eq:stringtrafo} (see Figure \ref{fig:enhancementpoints}).
For example at the $\mathbb{Z}_4$ point the two string charges are interchanged (with an additional minus sign for one). 

Since this symmetry is gauged there is no preferred duality frame, i.e., one cannot distinguish whether a string comes from wrapping an M2-brane on one or the other 1-cycle of the internal $T^2$ until one chooses a basis (and in turn fixes the gauge). Note that monodromies involving duality transformations around them are gauge invariant and therefore physically meaningful.\footnote{These can be associated with the circle reduction of general $[p,q]$-7-branes in the present setup, see also \cite{Dierigl:2020lai}.} Similarly, domain walls between different duality frames have a physical meaning. This distinction also appears in the definition of marked moduli spaces, important in the discussion of \cite{Raman:2024fcv}.

\vspace{0.5cm}

The duality groups from which the discrete gauge symmetries emerge at special points in the moduli space are:
\begin{itemize}
    \item{$N=32$ supercharges: In the maximal supersymmetric cases the duality group is given by the U-duality groups $G^D_{\mathrm{U}}(\mathbb{Z})$.}
    \item{$N=16$ supercharges: We discuss discrete symmetries coming from the T-duality group. In $D$ dimensions it takes the form SO$(26-D,10-D;\mathbb{Z})$. We further restrict to situations where the discrete symmetries are not embedded in continuous gauge symmetries (for example as part of the Weyl group of non-Abelian gauge theories).}
    \item{$N=8$ supercharges: We mainly focus on theories in six dimensions, where the discrete symmetries appear as subgroups of the symmetries in the tensor multiplet sector. For $T$ tensor multiplets the duality group is SO$(1,T;\mathbb{Z})$. We also describe a strategy to bound discrete symmetries acting on the hypermultiplets assuming the form of the symmetry action being a permutation of hypers.}
\end{itemize}
We will focus on discrete Abelian subgroups of the duality symmetries above and find upper bounds for:
\begin{itemize}
    \item{Maximal prime order element generating $\mathbb{Z}_p$ (also maximal prime powers $\mathbb{Z}_{p^r}$)}
    \item{Maximal order element generating $\mathbb{Z}_N$}
\end{itemize}
The first bound naturally enters the discussion of the second, since cyclic groups decompose according to their prime factors
\begin{equation}
    \mathbb{Z}_N = \mathbb{Z}_{p_1^{s_1}} \oplus \dots \oplus \mathbb{Z}_{p_\ell^{s_\ell}} \,,
\end{equation}
with $p_1^{s_1} \cdot ... \cdot p_\ell^{r_\ell} = N$. 
Beyond these questions, for some examples we also investigate at which special points in the moduli space these symmetries are realized.

\section{32 supercharges}
\label{sec:32}

We begin by analyzing theories with maximal supersymmetry, i.e., $32$ real supercharges, in $D$ dimensions. These theories possess a large non-Abelian discrete gauge symmetry given by the U-duality group $G^D_{\mathrm{U}} (\mathbb{Z}) $, see for example \cite{Cremmer:1978ds, Cremmer:1979up, Julia:1980gr, Cremmer:1980gs, Hull:1994ys, Obers:1998fb}. At generic points in the moduli space, $\mathcal{M}_{\mathrm{U}}$, this U-duality group is broken spontaneously by the vacuum expectation value of the scalar fields in the theory. But there are special points in the moduli space of the theory where subgroups of the full U-duality are restored. Here, we are interested in investigating the maximal cyclic subgroups of $G^D_{\mathrm{U}} (\mathbb{Z})$. 

In dimension $D = 11 - k$ the U-duality groups can be understood as the group generated by the two non-commuting subgroups (see \cite{Obers:1998fb})\footnote{We restrict the attention to the bosonic U-duality group. The inclusion of fermions requires a non-trivial $\mathbb{Z}_2$ extension, \cite{Pantev:2016nze, Chakrabhavi:2025bfi}. If this extension also affects the $\mathbb{Z}_N$ subgroup, it can enhance to $\mathbb{Z}_{2N}$ but will not lead to parametrically larger discrete groups.}
\begin{equation}
G^D_{\mathrm{U}} (\mathbb{Z}) = \text{SL}(k;\mathbb{Z}) \bowtie \text{Spin}(k-1,k-1;\mathbb{Z}) \,.
\label{eq:bowtie}
\end{equation}
The origin of these groups becomes apparent when describing maximal supergravity as torus compactifications of 11d supergravity. The SL$(k;\mathbb{Z})$ is the group of large diffeomorphisms of $T^k$ and the Spin$(k-1,k-1;\mathbb{Z})$ is the T-duality group of the type II theories on $T^{k-1}$. This procedure leads to the following U-duality groups:
\begin{equation}
\begin{array}{ c | c | c }
D & G^D_\mathrm{U} (\mathbb{Z}) & K^D_{\mathrm{U}} \\ \hline 
9 & \text{SL}(2;\mathbb{Z}) & \mathrm{SO}(2) \\ 
8 & \text{SL}(3;\mathbb{Z}) \times \text{SL}(2;\mathbb{Z}) & \mathrm{SO}(3) \times \mathrm{SO}(2) \\
7 & \text{SL}(5;\mathbb{Z}) & \mathrm{SO}(5) \\
6 & \text{Spin}(5,5;\mathbb{Z}) & \big( \mathrm{Spin}(5) \times \mathrm{Spin}(5) \big) / \mathbb{Z}_2\\
5 & \mathrm{E}_{6(6)}(\mathbb{Z}) & \mathrm{USp}(8) / \mathbb{Z}_2 \\
4 & \mathrm{E}_{7(7)}(\mathbb{Z}) & \mathrm{SU}(8) / \mathbb{Z}_2 \\
3 & \mathrm{E}_{8(8)}(\mathbb{Z}) & \mathrm{Spin}(16) / \mathbb{Z}_2
\end{array}
\label{eq:Ugroups}
\end{equation}
Here, we also denoted the maximal compact subgroup $K^D_{\mathrm{U}}$ of the continuous U-duality groups $G^D_{\mathrm{U}}(\mathbb{R})$, which will be useful in the derivation of the maximal order elements. The moduli spaces are given by
\begin{equation}
    \mathcal{M}_{\mathrm{U}} = \frac{G^D_{\mathrm{U}} (\mathbb{R})}{G^D_{\mathrm{U}}(\mathbb{Z}) \times K^D_{\mathrm{U}}} \,,
\end{equation}
parameterized by the scalar fields in the gravity multiplet.

For SL$(2;\mathbb{Z})$, as discussed in the example above, these properties are straightforward, and the biggest prime order is given by $p = 3$ associated to the $\mathbb{Z}_3$ matrix in \eqref{eq:SL2matrices}, the biggest prime power is given by $4 = 2^2$ and also given in \eqref{eq:SL2matrices}. The maximal cyclic subgroup is $\mathbb{Z}_6$ and associated to the element
\begin{equation}
\mathbb{Z}_6: \quad \begin{pmatrix} 1 & 1 \\ -1 & 0 \end{pmatrix} \in \mathrm{SL}(2;\mathbb{Z}) \,.
\end{equation}
These symmetries are realized at the special points in moduli space $\tau = e^{2 \pi i /3}$ and $\tau = i$, see \eqref{eq:SL2matrices}, which are invariant under $\mathbb{Z}_N$. 

In the following we will derive the bounds and their realizations for the remaining cases in dimensions $D < 9$ with $32$ supercharges.

\subsection{Bounds from U-duality}
\label{subsec:Ubounds}

For $D =8$ we gain an SL$(3;\mathbb{Z})$ factor, which however has the same maximal order elements as SL$(2;\mathbb{Z})$. We therefore find the same bounds as above, namely
\begin{equation}
    \mathbb{Z}_p = \mathbb{Z}_3 \,, \quad \mathbb{Z}_{p^r} = \mathbb{Z}_4 \,, \quad \mathbb{Z}_N = \mathbb{Z}_6 \,,
\end{equation}
where it is implied that these are the maximal values. All of these have realizations as embedding in the SL$(3;\mathbb{Z})$ and SL$(2;\mathbb{Z})$ factor, respectively.

For $D = 7$ one has the U-duality group SL$(5;\mathbb{Z})$ and again one can use the general bounds from GL$(5;\mathbb{Z})$, which are the same as for GL$(4;\mathbb{Z})$, giving:
\begin{equation}
    \mathbb{Z}_p = \mathbb{Z}_5 \,, \quad \mathbb{Z}_{p^r} = \mathbb{Z}_8 \,, \quad \mathbb{Z}_N = \mathbb{Z}_{12} \,,
\end{equation}
as maximal cyclic subgroups.

Things become a little more involved for $D=6$ with U-duality group Spin$(5,5;\mathbb{Z})$. Nevertheless, we can use the approach of quasicrystals, see \cite{Harvey:1987da, Baykara:2024vss}, to see that for SO$(5,5;\mathbb{Z})$ the bounds follow from GL$(8;\mathbb{Z})$. Even taking into account the non-trivial $\mathbb{Z}_2$ extension to Spin$(5,5;\mathbb{Z})$ these bounds do not change\footnote{Basically the GL$(8;\mathbb{Z})$ elements are decomposed into $2 \times 2$ rotation matrices, which produce factors of $(-1)$ for $2\pi$ rotations on fermions; for both the $\mathbb{Z}_{16}$ and $\mathbb{Z}_{60}$ the number of minus signs is even and there is no enhancement. (A different situation for example would be given by a $\mathbb{Z}_2$ element $\mathrm{diag}(-1,-1,1,\dots,1)$ in SO$(5,5;\mathbb{Z})$, whose square acts as $(-1)^F$ in Spin$(5,5;\mathbb{Z})$ and there is enhancement.)}
\begin{equation}\label{eq:32-Zp}
    \mathbb{Z}_p = \mathbb{Z}_7 \,, \quad \mathbb{Z}_{p^r} = \mathbb{Z}_{16} \,, \quad \mathbb{Z}_N = \mathbb{Z}_{60} \,.
\end{equation}
Note that GL$(8;\mathbb{Z})$ and not GL$(10;\mathbb{Z})$ appears since one demands a decomposition into rotations acting in the fixed signature parts, requiring an even rank (for left- and right-movers).

Finally, for $D \in \{ 5 \,, 4 \,, 3\}$ the U-dualities are given by exceptional groups of split real form E$_{k(k)}(\mathbb{Z})$. One can obtain naive upper bounds by considering GL$(\mathrm{dim}(\mathbf{R}); \mathbb{Z})$, with $\mathrm{dim}(\mathbf{R})$ the dimension of the defining representation, for the maximal order elements this would yield, see \cite{LEVITT1998630},
\begin{equation}
\begin{split}
    \mathrm{E}_{6(6)}(\mathbb{Z}):& \quad \mathbb{Z}_N \subset\mathbb{Z}_{9240} \subset \mathrm{GL}(27;\mathbb{Z}) \,, \\
    \mathrm{E}_{7(7)}(\mathbb{Z}):& \quad \mathbb{Z}_N \subset \mathbb{Z}_{3063060} \subset \mathrm{GL}(56;\mathbb{Z}) \,, \\
    \mathrm{E}_{8(8)}(\mathbb{Z}):& \quad \mathbb{Z}_N \subset \mathbb{Z}_{21906018973959120} \subset\mathrm{GL}(248;\mathbb{Z}) \,.
\end{split}
\label{eq:naiveUbounds}
\end{equation}
Clearly, these bounds are not very satisfying which is why we employ a different strategy to extract the maximal order elements. For that, we first note that since $\mathbb{Z}_N$ is compact it should be associated to an element in the maximal compact subgroup $K^D_{\mathrm{U}}$ of the U-duality group. The question we need to answer now is how an element in $K^D_{\mathrm{U}}$ embeds into $G^D_{\mathrm{U}}(\mathbb{R})$ and whether it can be transformed into an integer matrix via a change of basis. For the second part we again require that the element can be brought to a form involving the companion matrix of cyclotomic polynomials as described in Section \ref{subsec:rough}. In particular, the eigenvalues have to be of the form $e^{2 \pi i r/m}$ of the involved factors $m$ and ``totatives'' $r$ (i.e., integers $<m$ that are coprime to $m$). Note, however, that this is a necessary and not sufficient criterion, since the lift of the element $K^D_{\mathrm{U}}$ might not be conjugate to an integer element in $G^D_{\mathrm{U}}(\mathbb{Z})$. To prove a realization of the bound we can for example construct a specific element of order $N$. For the first part we realize that the maximal compact subgroup acts in a certain representation given by\footnote{That they cannot act in the defining representations can already be seen from the global group structure of the U-duality group as indicated in Table \eqref{eq:Ugroups}.}
\begin{equation}
\begin{split}
    \mathrm{USp}(8)/\mathbb{Z}_2:& \enspace \mathbf{27} \quad \text{anti-symmetric} \,, \\
    \mathrm{SU(8)}/\mathbb{Z}_2:& \enspace \mathbf{28} \oplus \overline{\mathbf{28}} \quad \text{anti-symmetric and its conjugate} \,, \\
    \mathrm{Spin}(16)/\mathbb{Z}_2:& \enspace \mathbf{120} \oplus \mathbf{128} \quad \text{anti-symmetric and spinor} \,,
\end{split}
\label{eq:compactrep}
\end{equation}
from which we can construct the eigenvalues of the corresponding element in E$_{k(k)} (\mathbb{R})$.

For $D = 5$ the eigenvalues of an element in the fundamental representation of USp$(8)$ come in complex conjugate pairs. Since any element of a compact, connected Lie group can be conjugated to an element in the maximal torus we can write it as the diagonal
\begin{equation}
    \text{diag}(\lambda_1 \,, \overline{\lambda}_1 \,, \dots \,, \lambda_4 \,, \overline{\lambda}_4) \,,
\end{equation}
with $|\lambda_i| = 1$. The same element in the anti-symmetric representation is also diagonal with eigenvalues given by the product of two different eigenvalues (there is an invariant 1-dimensional subspace determined by the Sp-invariant pairing, reducing the dimension of the representation from 28 to 27). This is how it acts on the $\mathbf{27}$ in the U-duality group. As shown in Appendix \ref{app:compact} the maximal order elements follow from the element in the maximal compact subgroup, leading to the bound
\begin{equation}
    \mathbb{Z}_p = \mathbb{Z}_{13} \,, \quad \mathbb{Z}_{p^r} = \mathbb{Z}_{32} \,, \quad \mathbb{Z}_N = \mathbb{Z}_{60} \,,
\end{equation}
which improves significantly on \eqref{eq:naiveUbounds}.

For $D = 4$ an element in the fundamental of SU$(8)$ can be diagonalized to
\begin{equation}
    \mathrm{diag}(\lambda_1 \,, \lambda_2 \,, \dots, \lambda_8 ) \,, \quad (\lambda_1 \lambda_2 \dots \lambda_8) = 1 \,,
\end{equation}
with $|\lambda_i| = 1$. The same element in the anti-symmetric representation is also diagonal with eigenvalues given by products $\lambda_i \lambda_j$, with $i \neq j$. Including the complex conjugate eigenvalues, this can naturally be understood as an element in GL$(56;\mathbb{R})$. For this to be transformable to an element in GL$(56;\mathbb{Z})$ with integer entries, we obtain the bounds
\begin{equation}
    \mathbb{Z}_p = \mathbb{Z}_{19} \,, \quad \mathbb{Z}_{p^r} = \mathbb{Z}_{64} \,, \quad \mathbb{Z}_N = \mathbb{Z}_{240} \,,
\end{equation}
where the maximal order depends on the assumption of a coprime decomposition and might not be sharp, see Appendix \ref{app:compact}. 

Finally, for $D = 3$ we need to further refine our arguments since an element in Spin$(16)$ acts in the composite representation $\mathbf{120} \oplus \mathbf{128}$. In Appendix \ref{app:compact} we derive bounds from the antisymmetric part resulting in
\begin{equation}
\mathbb{Z}_{p} = \mathbb{Z}_{31} \,, \quad  \mathbb{Z}_{p^r} = \mathbb{Z}_{256} \,, \quad \mathbb{Z}_N = \mathbb{Z}_{6930} \,.
\end{equation}
Here, the prime order bound is exact while both the prime power as well as the maximal order bounds (that depends on the coprime assumption in Appendix \ref{app:compact}) are likely not sharp. Nevertheless, they again represent a huge improvement compared to \eqref{eq:naiveUbounds}.

\subsection{Symmetric points}
\label{subsec:sympoints}

One might wonder whether these bounds are sharp, i.e. if these symmetries can be realized at special points in moduli space. This is in fact the case for every finite order element of the duality group for moduli spaces with non-positive sectional curvature such as $\mathcal{M}_\mathrm{U}$, as has been shown in \cite{bridson2013metric} (Corollary~2.8 on page 179). Basically, by the curvature property, there is a smallest sphere containing any orbit of the finite order element; the center of this sphere is a fixed point. For some bounds above we can find particular points in moduli space interpreted as torus compactifications of maximal supergravity theories. We will only do so for $\mathbb{Z}_p$ and $\mathbb{Z}_N$ in the following.

In $D = 9$ and $D = 8$ we can consider 11d supergravity on $T^2$ and $T^3$, respectively, and fix the complex structure of a $T^2$ (sub)torus to the $\mathbb{Z}_3$ and $\mathbb{Z}_6$ invariant point $\tau = e^{2 \pi i/3}$.

For $D = 7$ we can describe the $\mathbb{Z}_5$ as well as the $\mathbb{Z}_{12}$ element as an element of SL$(4;\mathbb{Z})$ acting as large diffeomorphism on $T^4$. The $\mathbb{Z}_5$ is restored when the lattice $\Lambda$ defining the 4-torus via $T^4 = \mathbb{R}^4/\Lambda$ is given by the root lattice of SU$(5)$, which contains an order 5 element in its Weyl group.\footnote{Note that since we are talking about the dimensional reduction of the maximal supersymmetric theories, there is no gauge enhancements at these points.} For the $\mathbb{Z}_{12}$ symmetric point we decompose $T^4$ into $T^2 \times T^2$ at the $\mathbb{Z}_3$ and $\mathbb{Z}_4$ symmetric points.

For $D = 6$ both the maximal order elements can be understood as elements of the T-duality group Spin$(4,4;\mathbb{Z})$. The different action on the left- and right-moving parts indicate that it is a quasicrystalline, and therefore non-geometric action that realizes the maximal discrete symmetry.

For $D=5$, one naturally inherits the maximal order elements of the six-dimensional theories, which once more are realized by a non-geometric action. This shows that the $\mathbb{Z}_{60}$ symmetry can indeed be realized. The realization of the $\mathbb{Z}_{13}$ is suggested by the work of \cite{steinberg1968lectures, Serre_Robinson_MacLachlan_Wall_1979, 2010arXiv1011.0346S}, since it cannot be embedded in the group of large diffeomorphisms or T-dualities it necessarily realizes a non-geometric action. 

In $D = 4$ the maximal prime order is given by $p = 19$ which again is realized by non-geometric action and guaranteed to exist by the work of \cite{steinberg1968lectures, Serre_Robinson_MacLachlan_Wall_1979, 2010arXiv1011.0346S}. For the maximal order we can realize an element of order $N =210$ by embedding in the U-duality group $\mathrm{SO}(6,6;\mathbb{Z})$, while we do not have an explicit construction of an element of order $N = 240$, which might therefore not describe a sharp bound. 

Finally, for $D=3$, even the maximal prime order element needs to combine both T-dualities and large diffeomorphisms and its existence is guaranteed by \cite{steinberg1968lectures, Serre_Robinson_MacLachlan_Wall_1979, 2010arXiv1011.0346S}. By combining T-duality transformations of order $210$, as for $D=4$ and a $\mathbb{Z}_4$ isometry of the axio-dilaton, we can realize an order $420$ element, and we do not expect the bound $\mathbb{Z}_{6930}$ to be sharp.

\section{16 supercharges}
\label{sec:16}

For maximal-rank theories with 16 supercharges, the duality group is $\mathrm{SO}(26-D,10-D;\mathbb{Z})$. A maximal order compact element is built from rotations in 2-planes entirely within the $26-D$ positive directions or within the $10-D$ negative directions, since a mixed-signature plane would give a boost. The total number of such rotation planes is $\lfloor (26-D)/2 \rfloor + \lfloor (10-D)/2 \rfloor$. When $D$ is odd, each block has one leftover 1-dimensional direction, and increasing $D$ to $D+1$ removes exactly these two unused directions, leaving the count of rotation planes unchanged. A similar phenomenon was noted for $D=5$ for $32$ supercharges, see below \eqref{eq:32-Zp}.

The duality group SO$(26-D, 10-D;\mathbb{Z})$ can be realized as the T-duality group of the heterotic string theories on a torus of dimension $(10-D)$. This connection to the heterotic string also suggests that one has to be careful to disentangle the genuinely discrete symmetries we are interested in from discrete subgroups of continuous gauge symmetries.

To illustrate this complication we consider the simple example of a $\mathbb{Z}_5$ symmetry. Using the discussion of Section \ref{subsec:rough} an element of this order can be realized within SL$(4;\mathbb{Z})$ and by the results of \cite{Baykara:2024vss} has a realization as an element of SO$(p,q;\mathbb{Z})$, for $p+q =4$ and $p$ and $q$ even. If the $\mathbb{Z}_5$ acts in a SO$(4,0;\mathbb{Z})$ inside the larger duality group, say $\mathrm{SO}(20,4;\mathbb Z)$, it might be an element of the Weyl group of an enhanced SU$(5)$ gauge symmetry. This happens if the geometric moduli are fixed to contain the associated root lattice and Wilson line as well as $B$-field moduli are switched off. To avoid this we can demand that the element embeds non-trivially in both signature subspaces, for example as an element of $\mathbb{Z}_5 \subset \mathrm{SO}(2,2;\mathbb{Z})$, then there is no point in moduli space where this becomes part of the Weyl symmetry of a non-Abelian gauge factor.

In fact, such maximal $\mathbb Z_N$ elements decompose into many small prime factors $p_i^{s_i}$ that are usually the symmetries of only the bosonic side.\footnote{For example, $p_i=2,3,5$ with low $s_i$ are easily identified to be in the Weyl group of $\mathrm{E}_8\times \mathrm{E}_8$ or $\mathrm{SO}(32)$.} Therefore, the bounds on $N$ we present here are not sharp, as they are most likely only realized as the symmetries of a Narain lattice $\Gamma^{26-D, 10-D}$ with roots, coming from continuous gauge symmetries.

For $D<9$, the largest possible prime (or prime-power) order symmetry in $\mathrm{SO}(26-D,10-D;\mathbb{Z})$ is determined by the cyclotomic degree. Writing $M=\phi(p)$ or $M=\phi(p^r)$, the allowed orders are those for which
\begin{subequations}\label{eq:het-ineq}
\begin{align}
M \le 36-2D,& \qquad D\ \text{even} \,,\\
M \le 34-2D,& \qquad D\ \text{odd} \,.
\end{align}
\end{subequations}
Thus the largest $\mathbb Z_{p}$ or $\mathbb Z_{p^r}$ symmetry is obtained by choosing the largest $p$ or $p^r$ satisfying \eqref{eq:het-ineq}.

When equality holds in \eqref{eq:het-ineq}, the corresponding maximal cyclotomic block is realized only on a non-unimodular lattice \cite[Appendix B]{Baykara:2024vss}, and therefore cannot embed in the even unimodular Narain lattice $\Gamma^{26-D,10-D}$. Hence in these cases the supergravity upper bound is potentially unsaturated.

Now consider strict inequality in \eqref{eq:het-ineq}, so that $M<36-2D$ (or $M<34-2D$ for odd $D$). Let $\Lambda^{a,b}$ be a non-unimodular quasicrystal of signature $(a,b)$ with
\begin{align}
a+b=M, \qquad a,b\in 2\mathbb{Z} \ge 0 \,,
\end{align}
carrying the relevant $\mathbb Z_{p}$ or $\mathbb Z_{p^r}$ symmetry. Since $\Lambda^{a,b}$ is irrational, it contains no roots, so the symmetry does not arise from a continuous gauge group.

If $\Lambda^{a,b}$ can be constructed such that the induced action on its discriminant group $\mathcal{D}_{\Lambda}$ is trivial, then any primitive embedding
\begin{align}
\Lambda^{a,b} \hookrightarrow \Gamma^{26-D,10-D} \,,
\end{align}
extends the $\mathbb Z_{p^r}$ symmetry to the full Narain lattice. By Nikulin’s embedding theorem \cite[Cor.~1.12.3]{VNikulin_1980}, a primitive embedding exists provided
\begin{align}
26-D \ge a, \qquad 10-D \ge b, \qquad (36-2D)-(a+b) > \ell(\mathcal{D}_{\Lambda}) \,,
\end{align}
where $\ell(\mathcal{D}_{\Lambda})$ is the minimal number of generators of the discriminant group.

Therefore, if \eqref{eq:het-ineq} holds with strict inequality and the symmetry preserves the discriminant group, then the resulting $\mathbb Z_{p}$ or $\mathbb Z_{p^r}$ symmetry is realized in heterotic string theory and saturates the supergravity bound.

The case of $D = 9$ is more subtle. The largest $p$ satisfying $\phi(p)=16$ is $p=17$. The $\mathbb{Z}_{17}$ is realized by constructing a non-unimodular $\Lambda^{16,0}$ with symmetry $\mathbb Z_{17}$ and gluing it to an appropriate $\Lambda^{1,1}$. However, $\Lambda^{16,0}$ is not a quasicrystal. In fact, it can be constructed as the root lattice of $\mathrm{A}_{16}$. Therefore the $\mathbb Z_{17}$ is realized as a Weyl symmetry. We cannot rule out the existence of another construction in which $\mathbb Z_{17}$ does not act on a root lattice. For example, the Conway subgroup symmetric compactifications \cite{Harvey:2017xdt,Baykara:2021ger} are another family in addition to quasicrystalline points that have large symmetries that do not come from continuous gauge groups. However, it turns out that there is no such $\Gamma^{17,1}$ construction for $D=9$, so the bound still remains unsaturated. 

The above approach produces the following upper bounds:
\begin{equation}\label{eq:16Q-bounds}
    \begin{array}{c | c | c | c }
        D & \mathbb{Z}_p & \mathbb{Z}_{p^r} & \mathbb{Z}_N \\ \hline
        9 & \mathbb{Z}_{17} & \mathbb{Z}_{32} & \mathbb{Z}_{840} \\
        8 & \mathbb{Z}_{19} & \mathbb{Z}_{32} & \mathbb{Z}_{2520} \\
        7 & \mathbb{Z}_{19} & \mathbb{Z}_{32} & \mathbb{Z}_{2520} \\
        6 & \mathbb{Z}_{23} & \mathbb{Z}_{32} & \mathbb{Z}_{5040} \\
        5 & \mathbb{Z}_{23} & \mathbb{Z}_{32} & \mathbb{Z}_{5040}  \\
        4 & \mathbb{Z}_{23} & \mathbb{Z}_{32} & \mathbb{Z}_{13860}  \\
        3 & \mathbb{Z}_{23} & \mathbb{Z}_{32} & \mathbb{Z}_{13860} \\ 
    \end{array}
\end{equation}
The bounds we have in \eqref{eq:16Q-bounds} are populated as follows: the bounds on $N$ come from \eqref{eq:GLmax} for $N=36-2D$ for even $D$ and $N=34-2D$ for odd $D$. We expect that they are realized only within continuous gauge groups so these bounds are unsaturated. The bounds on $p$ and $p^r$ come from the largest prime and prime power quasicrystalline action in $36-2D$ dimensions for even $D$ and $34-2D$ dimensions for odd $D$, which can be read from Table 11 in \cite{Baykara:2024vss}. We suspect that the construction outlined above saturates these (prime order) bounds, though we do not yet have a proof that it can always be carried out.

Since the main limiting factor to realize the maximal order element is the rank of the duality matrix, the bounds above also apply in the reduced rank cases, descendant for example from the CHL string.

\subsection{Realizations and symmetric points} 

For the maximal order elements, determining a maximal subgroup $\mathbb{Z}_N$, we can only define upper bounds with one of the complications being the embedding of the discrete groups within continuous gauge symmetry sectors. In the following we explore explicit realizations given by quasicrystalline compactifications, which act on even, self-dual lattices, where no group element acts only on the bosonic lattice (and might therefore be associated to non-Abelian gauge enhancement). For that we use Table 11 in \cite{Baykara:2024vss} and realize that we can use $\tfrac{1}{2}(10-D)$ independent quasicrystalline group actions, which corresponds to an orthogonal decomposition of the underlying lattice $\Gamma^{26-D,10-D}$. Maximizing the overall order $N$ of the symmetry action leads to;
\begin{equation}
    \begin{array}{c | c | c}
    D & \mathbb{Z}_N & \Gamma^{26-D,10-D} \\ \hline
    8 & \mathbb{Z}_{66} & (\Gamma^{18,2})_{\mathbb{Z}_{66}} \\
    7 & \mathbb{Z}_{66} & (\Gamma^{18,2})_{\mathbb{Z}_{66}} \oplus \Gamma^{1,1} \\
    6 & \mathbb{Z}_{252} & (\Gamma^{10,2})_{\mathbb{Z}_{21}} \oplus (\Gamma^{10,2})_{\mathbb{Z}_{36}} \\
    5 & \mathbb{Z}_{252} & (\Gamma^{10,2})_{\mathbb{Z}_{21}} \oplus (\Gamma^{10,2})_{\mathbb{Z}_{36}} \oplus \Gamma^{1,1} \\ 
    4 & \mathbb{Z}_{840} & (\Gamma^{12,4})_{\mathbb{Z}_{40}} \oplus (\Gamma^{10,2})_{\mathbb{Z}_{21}} \\
     & & (\Gamma^{10,2})_{\mathbb{Z}_{21}} \oplus (\Gamma^{6,2})_{\mathbb{Z}_{15}} \oplus (\Gamma^{6,2})_{\mathbb{Z}_{24}} \\
    3 & \mathbb{Z}_{840} & (\Gamma^{12,4})_{\mathbb{Z}_{40}} \oplus (\Gamma^{10,2})_{\mathbb{Z}_{21}} \oplus \Gamma^{1,1} \\
     & & (\Gamma^{10,2})_{\mathbb{Z}_{21}} \oplus (\Gamma^{6,2})_{\mathbb{Z}_{15}} \oplus (\Gamma^{6,2})_{\mathbb{Z}_{24}} \oplus \Gamma^{1,1}
    \end{array}
\end{equation}
where we also denote in how many orthogonal components the lattice decomposes and what order acts on the individual pieces (in one particular realization). As expected, reducing the amount of supersymmetry allows for the realization of larger discrete groups (as compared to Section~\ref{sec:32}), but the bounds in \eqref{eq:16Q-bounds} cannot be saturated using these constructions.

As for most of the symmetric points in the case of $32$ supercharges a (non-geometric) quasicrystalline action, which acts differently on the left- and right-movers on the string worldsheet, is required in order to realize these large discrete symmetries.

\section{8 supercharges}
\label{sec:8}

For eight real supercharges we are necessarily in dimensions $D \leq 6$ and we will focus on $D = 6$ in the following. There are several sources for discrete symmetries in $\mathcal{N} = (1,0)$ supergravity theories. We will mainly restrict discrete symmetries acting on the scalars of the tensor multiplets, which is closely related to the discussion above. For hypers we make the assumption (that we do not know to be the most general) that the symmetry acts as permuting hypermultiplet fields. To obtain these bounds we further improve the existing supergravity bounds on the rank of the gauge group and the number of neutral hypermultiplets, which might prove useful also in other context. 

Another source of discrete symmetries arises from discrete charges of individual hypermultiplets under continuous gauge symmetries which get Higgsed. A typical example is $\mathbb{Z}_N$ symmetries from a supersymmetric version of the Higgs mechanism that breaks U$(1)$ symmetries by condensing charge $N$ fields. We postpone a detailed study of this latter class to future work and only provide a brief argument using discrete anomaly constraints.

For $T$ tensor fields the supergravity theory defines an integral and unimodular (not necessarily even) lattice $\Lambda$ of signature $(1,T)$, which can be understood as the string charge lattice. We are interested in $\mathbb{Z}_N$ isometric automorphisms acting on this lattice, which then for special values of the moduli fields acts as a $\mathbb{Z}_N$ gauge symmetry in the supergravity theory. The richer structure of the moduli space with 8 supercharges introduces additional challenges, which we have not encountered in the higher supersymmetric cases above. In particular, it is not enough to find a self-dual lattice with signature $(1,T)$ that admits for an automorphism of order $N$, we further need to make sure that this lattice is realized as a point in the moduli space, as the tensor moduli space is bounded by loci leading to SCFT's. These complications will in general not allow us to obtain a sharp upper bound, but nevertheless we are able to derive various general upper bounds.

\subsection{General upper bounds on the tensor sector}

Since the transformations we are interested in act on the tensor field moduli, which determine the tension of BPS strings, they act on the string charge lattice. They thus have a realization as elements in GL$(T+1;\mathbb{Z})$ for which we derived general upper bounds in Section \ref{subsec:rough}.

For the maximal prime order we obtain the bound 
\begin{equation}
    p \leq T+1 \,,
\end{equation}
which is therefore correlated with the number of tensor multiplets. Similarly we obtain a rough upper bound for the maximal order, by the maximal order element of GL$(T+1;\mathbb{Z})$, which can be found for example in \cite{LEVITT1998630}.

Together with the results in \cite{Kim:2024eoa}, and under the same assumptions, we find that the maximal orders are finite. Moreover, \cite{Kim:2024eoa} obtains an upper bound $T \leq 193$, which has a realization in string theory. Given this upper bound, we find
\begin{equation}
\begin{split}
    \mathbb{Z}_p :& \quad p \leq 193 \,, \\
    \mathbb{Z}_N :& \quad N \leq 89048857617720 = 2^3 \cdot 3^2 \cdot 5 \cdot 7 \cdot 11 \cdot 13 \cdot 17 \cdot 19 \cdot 23 \cdot 29 \cdot 31 \cdot 37 \,,
\end{split}
\end{equation}
where the bound on $\mathbb{Z}_N$ originates from the largest order element of GL$(194;\mathbb{Z})$. We will attempt to sharpen this in the following.

\subsection{Constraints from the {\it H}-string}
\label{sec:H-string}

Every 6d $(1,0)$ supergravity theory with $T\ge1$ contains a particular class of BPS strings known as ``{\it H-string}'' \cite{Kim:2024eoa}. These are either the critical heterotic string or little strings in a charge class $f$ satisfying $f^2=0$ and $b_0\cdot f=2$, where $b_0$ denotes the anomaly vector associated to the gravitational anomaly. Their low-energy worldsheet dynamics are described by two-dimensional $(0,4)$ SCFTs. Unitarity of these worldsheet SCFTs constrains the possible 6d gauge algebras that induce worldsheet current algebras with level $k_i=f\cdot b_i>0$ on the string where $b_i$ denotes the anomaly vector of the gauge factor. We call such algebras {\it external} gauge algebras $G_{\rm ext}$. For example, the total rank of external gauge algebras is bounded above by 20 \cite{Kim:2019vuc,Lee:2019skh}.

We will now prove that the rank of any such external gauge algebra is further bounded by 16, except when the 6d supergravity does not contain any little string in its BPS spectrum, namely, for $T=1$ with no gauge algebra supported on the tensor multiplet of charge $f$ (i.e., $b_i\neq f$ for all $b_i$), as well as for $T=0$ (which does not admit an {\it H}-string). Equivalently, whenever the {\it H}-string can degenerate into a little string, either as a collection of self-dual strings of local 6d SCFTs (when $T>1$) or as an instantonic string in a little string theory (when $T\ge1$ and the tensor multiplet of charge $f$ supports a gauge algebra), the rank of $G_{\rm ext}$ has a stronger upper bound 16. By contrast, for $T=1$ with no gauge algebra on $f$, where a perturbative heterotic string description applies, the bound still remains as 20.

To establish this, we first note that the (0,4) superconformal algebra of the worldsheet theory on non-critical strings in local 6d SCFTs is distinct from that of the critical heterotic {\it H}-string. The 6d heterotic string has central charges $(c_L,c_R)=(20,6)$. These follow from the central charge formula for supergravity strings \cite{Kim:2019vuc}:
\begin{align}
    c_L = 3Q^2 + 9Q\cdot b_0+2 \ , \quad c_R = 3Q^2+3Q\cdot b_0 \ .
\end{align}
for a string charge $Q$. This expression relies on the anomaly inflow in the presence of the string and the identification of the worldsheet R-symmetry with the subgroup $\mathrm{SU}(2)_R\subset \mathrm{SO}(4)$ of the Lorentz symmetry acting on the transverse $\mathbb{R}^4$. However, for strings with $Q^2<0$ that appear in local 6d SCFTs, this formula yields negative central charges (except a single E-string). This means that the worldsheet R-symmetry for such strings cannot be the $\mathrm{SU}(2)_R$. Indeed, as emphasized in \cite{Kim:2019vuc}, the IR worldsheet theories of 6d SCFT strings exhibit an emergent $\mathrm{SU}(2)_I$ symmetry, inherited from the $\mathrm{SU}(2)_I$ R-symmetry of the bulk 6d SCFTs, and this $\mathrm{SU}(2)_I$ symmetry serves as the R-symmetry of the $(0,4)$ superconformal algebra of the low-energy 2d CFT, and yields positive central charges as expected.

There is a complementary argument for this fact using the instanton viewpoint. Most 6d SCFT strings (aside from the E- and M-strings) are instantonic and their worldsheet moduli spaces are parameterized by bosonic zero modes from 6d hypermultiplets charged under the gauge algebra. The $\mathrm{SU}(2)_R\subset \mathrm{SO}(4)$ symmetry in fact acts on these hypers and thus on the associated bosonic zero modes. This implies that $\mathrm{SU}(2)_R$ cannot become the $(0,4)$ R-symmetry because the right-moving R-symmetry in the IR CFT cannot act on the scalar fields in the moduli space of vacua \cite{Witten:1997yu}. Thus the superconformal algebra on instantonic non-critical strings differs from that of the critical heterotic string which instead has a $(0,4)$ algebra with $\mathrm{SU}(2)_R$ R-symmetry. The IR R-symmetry of the instanton strings turns out to be the emergent $\mathrm{SU}(2)_I$, under which the moduli scalars are neutral.

From this perspective, when the {\it H}-string degenerates into 6d SCFT/instantonic strings, the low energy theory decomposes into a product of two types of 2d $(0,4)$ SCFTs that intersect only pointwise in their (classical) moduli spaces. The first type corresponds to the heterotic string having a $(0,4)$ superconformal algebra with $\mathrm{SU}(2)_R$ R-symmetry. This $(0,4)$ SCFT does not see the $\mathrm{SU}(2)_I$ symmetry. Then at special loci in its moduli space where the {\it H}-string degenerates, there appears a second type of $(0,4)$ SCFT, whose R-symmetry is the emergent $\mathrm{SU}(2)_I$ symmetry, describing the low energy theory for 6d SCFT/instantonic strings. There can be multiple such second-type SCFTs when more than one LST lies in the same \emph{H}-string charge class.

We now claim that the external gauge algebra $G_{\rm ext}$ can couple only to non-supersymmetric chiral sectors in these two types of 2d SCFTs.
The reason is that the current algebra for $G_{\rm ext}$ must be shared by both the critical heterotic string and one of the non-critical strings into which the {\it H}-string can degenerate. The supersymmetric sectors (including moduli scalars), however, host distinct superconformal algebras and thus cannot support a common current algebra.

As explained above, the 2d $(0,4)$ SCFT on the heterotic string of the {\it H}-string has a superconformal algebra with the $\mathrm{SU}(2)_R$ R-symmetry and it possesses a hyper-K\"ahler moduli space. Since the heterotic string has $c_R=6$, the supersymmetric sector containing the hyper-K\"ahler moduli space  must be realized by a (twisted) hypermultiplet, which contributes 4 to the left-moving central charge.\footnote{For the 6d F-theory models, the moduli space described by this hypermultiplet is the dual K3 surface in the heterotic dual. It was also noted in \cite{Lee:2019skh} that, provided this hypermultiplet does not enter the current algebra, the rank of Abelian gauge algebras, which are also external gauge algebras in our terminology, is bounded by 16 for F-theory models.} It then follows that the left-moving central charge available to chiral current algebras from $G_{\rm ext}$ is at most $16=c_L-4$. Hence, whenever the 6d theory contains little strings, the total rank of the external gauge algebra is bounded by 16 (rather than 20).

The simplest examples arise when the 6d gravity theory contains a little string theory obtained from two intersecting E-string theories. In this situation, one may take the \emph{H}-string charge class $f=(1,1)$, where $(n,m)$ denotes the E-string charges. The external gauge algebra $G_{\rm ext}$  must then be a subgroup of $\mathrm{E}_8 \times \mathrm{E}_8$, where each $\mathrm{E}_8$ corresponds to the flavor symmetry of one E-string, as the positive-level condition $k>0$ with respect to the \emph{H}-string forces $G_{\rm ext}$ must intersect one or both E-string factors. Therefore, the rank of $G_{\rm ext}$ is bounded by 16.

\subsection{Refined upper bounds on tensor sector}

We now refine the general upper bound by exploiting the novel structure of 6d supergravity theories established in \cite{Kim:2024eoa}. It was shown there that the intersection patterns of tensor multiplets in any 6d $(1,0)$ supergravity theory can be realized by the topology of $\mathbb{P}^2$ or blow-ups of Hirzebruch surfaces $\mathbb{F}_n$, which are the allowed K\"aher bases of elliptic Calabi-Yau (CY) 3-folds. Two key properties from this investigation are: 
\begin{itemize}
    \item{Iterative blow-downs of tensors with self-intersection $-1$ ultimately brings the tensor intersection form to that of a Hirzebruch surface $\mathbb{F}_n$.}
    \item{Tensor multiplets that do not intersect an {\it H}-string must be embedded in a little string theory (LST) in the same charge class as the {\it H}-string.}
\end{itemize}
See Section 3.2 and 3.3 of \cite{Kim:2024eoa} for more details. Using these properties, we derive a refined upper bound as follows.

Since our focus is on the action on the tensor moduli space and on its maximal order, we can assume $T> 1$. Then the tensor charge lattice contains a {\it H}-string charge $f$, and the BPS cone, i.e., the cone of BPS string charges within the lattice, is generated by BPS generators $\{\mathcal{C}_i\}$ with $\mathcal{C}_i^2<0$, which are all classified in \cite{Heckman:2015bfa,Bhardwaj:2015xxa,Bhardwaj:2015oru,Bhardwaj:2019hhd}. We distinguish `external' generators satisfying $f \cdot \mathcal{C}_i>0$ for the \emph{H}-string charge $f$ from those with $f\cdot \mathcal{C}_i=0$. The latter must be embedded in an LST in the charge class $f$ \cite{Kim:2024eoa}. The discrete symmetries of the tensor moduli space are realized as subgroups of the symmetry group that permutes the BPS generators while preserving their intersection structure.

We will show that the maximal order of such discrete symmetries permuting the $\mathcal{C}_i$'s is bounded by 26. The argument proceeds as follows. First, the number of external generators supporting gauge algebras is bounded by 16 due to the bound {\rm rank}$(G_{\rm ext})\le 16$. This already lies within the proposed bound 26. Then the remaining allowed external generators are the E- and M-strings, whose gravitational anomaly contributions are $\Delta_i=29$ and 30, respectively. All the other generators satisfy $f\cdot \mathcal{C}_i=0$ and are contained in an LST of the charge class $f$.

Little string theories in the charge class $f$ have the P-type endpoint \cite{Kim:2024eoa,Bhardwaj:2015xxa}. The tensor intersections of this type form a tree-shaped diagram built from “nodes’’ and “links’’ (see \cite{Heckman:2015bfa} for definitions) without any loops. The elementary building blocks are given by an internal link together with one half of each adjacent node, as well as side links. The classification of nodes and links in \cite{Heckman:2015bfa} implies that each such block admits at most an order-4 permutation symmetry. Hence a single block cannot generate a large permutation group. Consequently, any large permutation symmetry must arise from arranging many identical blocks and permuting them collectively.

As noted in \cite{Kim:2024eoa}, each building block in an LST admitting a P-type endpoint contributes a positive gravitational anomaly $\Delta_i\ge 29$.\footnote{There is a single exception: the $\mathfrak{e}_8\oplus \mathfrak{so}_8$ conformal matter together with half-nodes of the neighboring $\mathfrak{e}_8$ and $\mathfrak{so}_8$, for which $\Delta_i=27$. However, embedding this blocks in an LST always requires additional blocks, and so the allowed number of this exceptional block in supergravity is strongly constrained and thus smaller relative to other cases.} As shown above, the external E- and M-strings also satisfy $\Delta_i\ge 29$. The gravitational anomaly constraint $H-V+29T=273$ therefore bounds the total number of these blocks together with the external E/M-strings. The maximum number of these components is then controlled by the number of vector multiplets for the external gauge algebras (which contributes negatively to the anomaly), whereas contributions from gauge algebras inside the LSTs are already accounted for when we compute those with $\Delta_i\ge 29$ of the building blocks. 

Now remember that, as we have shown earlier, the rank of external gauge algebra for $T>1$ is bounded by 16. This means  that the maximum number of external vectors is 496 from $\mathrm{E}_8\times \mathrm{E}_8$ or $\mathrm{SO}(32)$. Hence, the maximum number of identical components (or blocks) $n_{\rm max}$ that we can permute while leaving the tensor intersection structure fixed is 
\begin{align}
29\cdot n_{\rm max} \le 273+496 \quad\rightarrow \quad n_{\rm max} =26 \ .
\end{align} 
This shows that the maximal permutation group acting on the tensor moduli space is $S_{26}$, and the maximal prime order of a discrete symmetry is $23$. The maximal order $N$ is given by the least common multiple of a partition of $26$, here $7+5+9+4+1$, and is given by $N = 1260$. 

In F-theory, one can engineer a supergravity theory containing 24 identical building blocks with the same intersection structure. Its K\"ahler base is a Hirzebruch surface $\mathbb{F}_{12}$ with 24 blow-ups at special points, which leads to two $-12$ curves supporting $\mathrm{E}_8\times \mathrm{E}_8$ gauge symmetry and 24 LSTs, each formed by two E-strings intersecting one of the two $-12$ curves. This construction yields a theory with $T=25$ and $\mathrm{E}_8\times \mathrm{E}_8$ gauge symmetry. On the tensor charge lattice, there is an $S_{24}$ permutation symmetry that exchanges the 24 identical LSTs. However, the $S_{24}$ permutation group turns out to act also on the complex structure moduli of neutral hypermultiplets, and the symmetry on the charge lattice is consequently broken to a subgroup, say $\mathbb{Z}_{24}$. It's a nontrivial task to determine all the other possibilities.\footnote{An easier example $dP_8$ is understood in detail, where the maximal unbroken subgroup has order 144, way smaller than the full symmetry group and the Weyl group.}

This geometric example indicates that a purely field-theoretic analysis does not readily reveal the precise symmetry of the tensor moduli space. Any symmetry that also acts on complex structure moduli may not be manifest in the low-energy supergravity. Thus, even if one constructs a tensor charge lattice saturating the bound $n_{\rm max}$ by introducing 26 identical blocks, the actual symmetry acting on these blocks will generically be smaller. Therefore, the maximal permutation group $S_{26}$, as well as the bounds on $\mathbb{Z}_p$ and $\mathbb{Z}_N$;
\begin{equation}
    \mathbb{Z}_p = \mathbb{Z}_{23} \,, \quad \mathbb{Z}_N = \mathbb{Z}_{1260} \,,
\end{equation}
derived from it, should rather be viewed as an upper bound, and it may not be realized in any consistent gravity theory.

On the other hand, an F-theory construction with $T=24$ and $\mathrm{E}_8 \times \mathrm{E}_8$ gauge symmetry realizes the maximal prime order $\mathbb{Z}_p = \mathbb{Z}_{23}$. The associated elliptic 3-fold is constructed over the Hirzebruch base $\mathbb{F}_{12}$ with 23 blow-ups. Alternatively, the same geometry can be described as a 3-fold over $\mathbb{F}_{12}$ with 24 blow-ups, where one of the exceptional curves shrinks to zero size. Hence, the bound $p = 23$ is sharp.

\subsection{Hypermultiplet sector}
\label{sec:hyper}

We now turn to discrete symmetries acting on the hypermultiplet moduli space. In this work we mainly focus on discrete symmetries which act as permutations of hypermultiplets and leave a detailed study of the more general case to future work.

We will first prove that the number of neutral hypermultiplets, denoted $H_0$, in any 6d supergravity theory is bounded above by 492. This improves the previous bound $H_0\le1064$ obtained in \cite{Kim:2024eoa} and agrees with the Hodge number bound $h^{2,1}(CY_3)\le 491$ in geometry found in \cite{Taylor:2012dr}. Moreover, this bound is sharp and is saturated by an F-theory model on an elliptic Calabi–Yau 3-fold over the Hirzebruch base $\mathbb{F}_{12}$ with an $\mathrm{E}_8$ gauge algebra on the $-12$ curve, for which $H_0=492$.

The gravitational anomaly cancellation condition implies that in order to maximize the number of neutral hypermultiplets, we need to maximize the number of vector multiplets while keeping both the number of charged hypermultiplets and the number of associated tensor multiplets small. A gauge algebra $G_i$ must be supported on a tensor charge $b_i$ with either $b_i^2>0$ or $b_i^2\le 0$. For the former, the gauge anomaly cancellation condition {(see Appendix \ref{app:6danomaly}) tells us that, since $b_i^2>0$ demands much more hypers than vectors, the gravitational anomaly contribution $\Delta_i=H_i-V_i$ cannot be negative unless the gauge algebra $G_i$ has very large dimension and many of its charged hypermultiplets are also charged under other gauge algebras $G_j$ with $b_j^2<0$. However, because matter sectors with $b_j^2< 0$ do not furnish exceptional flavor symmetries, such multi-charged hypers cannot exist for the exceptional gauge algebras $G_i= \mathrm{F}_4, \mathrm{E}_{6,7,8}$. Also, classical $\mathrm{SU}, \mathrm{SO}, \mathrm{Sp}$-type gauge algebras with the bounded ${\rm rank}(G)\le 20$, as well as $\mathrm{G}_2$ algebra, supported on $b_i^2>0$ require far more charged hypers than their algebra dimensions, and thus their gravitational anomaly contributions eventually become positive even in the presence of multi-charged hypers. Therefore, introducing any gauge algebra with $b_i^2>0$ (and likewise on $b_i^2=0$) can only decrease the number of neutral hypermultiplets. This was also observed in \cite{Hamada:2023zol,Hamada:2024oap,Hamada:2025vga} in the classification program (when $\mathrm{U}(1), \mathrm{SU}(2), \mathrm{SU}(3)$ gauge algebras are absent), see also \cite{Birkar:2025rcg, Birkar:2025gvs} for a more geometric perspective.

We next analyze gauge algebras with $b_i^2<0$. For $T=0$, there is no tensor charge with $Q^2<0$, and thus the maximal number of neutral hypers is simply 273, realized by the pure supergravity theory with $V=0$. For $T=1$, the tensor base admits at most one negatively self-intersecting tensor \cite{Kim:2024eoa}. The choice that maximizes $H_0$ is an $\mathrm{E}_8$ gauge algebra with $b_i^2=-12$ and no charged hypers. This theory is precisely the F-theory model over the Hirzebruch surface $\mathbb{F}_{12}$, which saturates the bound $H_0\le 492$. Any additional matters to this setup only reduces $H_0$.

When $T>1$, such a gauge algebra can either lie within an LST component of the {\it H}-string charge class $f$ (when $b_i\cdot f=0$) or be external with $b_i\cdot f>0$. In the first case, as noted in \cite{Kim:2024eoa} and section \ref{sec:H-string}, the corresponding component, which is a collection of intersecting nodes and links, contributes positively to the gravitational anomaly, roughly $\Delta\ge29$.  Adding such a component decreases $H_0$. Thus, $H_0$ can be increased only by adding external gauge algebras, whose rank is bounded by $16$ as derived in Section~\ref{sec:H-string}. One might therefore attempt to use rank-16 external algebras of large dimension to exceed the proposed bound $H_0=492$. However, this is also forbidden due to constraints on the tensor multiplets and their intersections. As proven in \cite{Kim:2024eoa}, the tensor multiplet sector and its intersection structure in any 6d supergravity theory must always be realized via the topology of an admissible K\"ahler base for elliptic 3-folds. This means that any tensor intersection form with $T>1$ can be blown down to a one-dimensional subspace identical to that of a Hirzebruch surface.\footnote{Here, a {\it blow-down} denotes a projection orthogonal to a tensor multiplet with $Q^2=-1$ in the tensor moduli space; see Section~3.2 of \cite{Kim:2024eoa}.}

As an illustration, imagine a $T=2$ theory with an external $\mathrm{E}_8\times \mathrm{E}_8$ gauge algebra supported on two $(-12)$ tensors, which yields the gravitational anomaly $-V+29T=-438$ and thus $H_0=273+438=711$. However, two tensor charges with $Q^2=-12$ cannot be realized in any tensor intersection form with $T=2$ that can be blown down to that of a Hirzebruch surface. In fact, realizing two $(-12)$ tensors requires introducing at least 24 additional components with $\Delta\ge29$, which corresponds to at least 24 blow-ups from a Hirzebruch surface in geometry. So in this $\mathrm{E}_8\times \mathrm{E}_8$ case, the number of neutral hypers cannot exceed $73=273+496-29\times 24$. Similarly, $\mathrm{E}_8\times \mathrm{E}_7$ external gauge algebra requires at least 20 more components and $\mathrm{E}_7\times \mathrm{E}_7$ at least 16 more components, and so on. In all such cases including two or more external gauge algebras, $H_0$ cannot exceed 492.

One might then ask whether a single external rank-16 algebra can yield larger $H_0$. This is conceivable for external $\mathrm{SU}$, $\mathrm{SO}$, or $\mathrm{Sp}$ algebras with $b_i^2\ge-4$. However, these always require many charged hypermultiplets, so their net contribution to the gravitational anomaly becomes positive, except for $\mathrm{SO}(32)$ algebra with $b_i^2=-4$. Even in this $\mathrm{SO}(32)$ case, anomaly cancellation requires 24 fundamentals, and thus the most negative gravitational anomaly contribution (by assuming that all fundamentals are also charged under other gauge algebras) is $-112=-32\times 31/2+24\times 32/2$, which is still greater (less negative) than $-248$ for a single $\mathrm{E}_8$ on an external $(-12)$ tensor. Thus, no external $\mathrm{SU}, \mathrm{SO}, \mathrm{Sp}$ gauge algebra can can produce a sufficiently negative anomaly contribution to increase $H_0$.

Taken together, these results establish a universal bound on the number of neutral hypermultiplets,
\begin{align} 
    H_0\le 492 \,,
\end{align} 
for any 6d supergravity. This bound is sharp and is saturated by the F-theory model with $T=1$ and an $\mathrm{E}_8$ gauge symmetry on a $(-12)$ tensor multiplet.

Moreover, this result allows us to derive a bound on the permutation group acting on charged hypermultiplets as well. We claim that it is at most $S_{493}$, implying that no more than 493 hypermultiplets can carry the same charge. Our argument crucially relies on the bound $H_0\le492$ and proceeds as follows.

First, we can easily show that, for a gauge group $G_i$ with $b_i^2\le0$, the multiplicity of identically charged hypermultiplets cannot exceed 493. If the gauge group is external with respect to an {\it H}-string, the rank of $G_i$ is bounded by 20 (reduced to 16 in the presence of an LST). In this case, the 6d SCFT classification \cite{Heckman:2015bfa,Bhardwaj:2015xxa} shows that the maximal multiplicity in a fixed representation ${\bf R}$ is 48, which occurs for $G_i=\mathrm{Sp}(20)$ on a $-1$ tensor. If instead the tensor with charge $b_i$ sits inside an LST, the permutation group is a subgroup of the LST flavor symmetry. We then notice from  \cite{Kim:2024eoa} that the LST configuration achieving maximal rank 480 in any 6d supergravity is obtained by assuming all flavor symmetries in an LST are gauged. This implies that the maximal rank of flavor symmetry in an LST embedded in a supergravity is 480 (in practice, much smaller). Hence, for $b_i^2\le0$, the number of (full) hypermultiplets in an identical representation of $G_i$ is bounded by $480$.

Next, if the gauge algebra $G_i$ is supported on $b_i$ with $b_i^2>0$, it must be external and intersect every {\it H}-string, and thus ${\rm rank}(G_i)\le 20$. Under this rank bound, any gauge algebra $G_i$ admitting a large multiplicity (of order $\mathcal{O}(10^2)$) of identically charged hypermultiplets can be fully Higgsed. The Higgsing leaves some neutral hypermultiplets while the rest are eaten by vector pairs. This means that the maximal number of hypermultiplets in a fixed representation (or charge) {\bf R} is $\frac{H_0^{\rm max}+{\rm dim}(G_i)}{{\rm dim}({\bf r})}$. For example, for $G=\mathrm{SU}(2)$, this yields $(H_0^{\rm max}+3)/2 \simeq 247$, where $3={\rm dim}(\mathrm{SU}(2))$ and the factor $1/2$ accounts for fundamentals ${\bf R}={\bf 2}$. Since the rank is bounded by 20, the maximum is achieved for $G=\mathrm{U}(1)$ with a common charge for all hypermultiplets. In this case the bound is $H_0^{\rm max}+1=493$, though we do not know whether this value can be achieved as an anomaly-free configuration and is actually realized. 

In conclusion, we find that the bound on the number of identical hypermultiplets in 6d $(1,0)$ theories is 493. Hence the maximal permutation group is $S_{493}$ with the maximal prime order cyclic group $\mathbb{Z}_p = \mathbb{Z}_{491}$.

\subsection{Discrete anomalies}
\label{subsec:discanom}

Finally, let us briefly comment on a constraint for hypermultiplets charged under the discrete symmetries, i.e., not acting as a permutation on the hypermultiplets. While we are not able to produce a universal bound, we can describe a general strategy that can be applied after a specific model is chosen. This possibility to constrain symmetries under which hypermultiplets are charged relies on the absence of discrete gauge anomalies. In fact, as soon as there are charged hypermultiplets there will be discrete gauge anomalies classified by the bordism group $\Omega^{\mathrm{Spin}}_7 (B \mathbb{Z}_N)$, see, e.g., \cite{Garcia-Etxebarria:2018ajm, Hsieh:2018ifc, Monnier:2018nfs, Dierigl:2022zll, Dierigl:2025rfn, Cheng:2025ikd}, which is non-trivial. 

For $N = p$ prime ($p>3$), on which we will focus in the following, this group is given by \cite{Debray:2023yrs,Dierigl:2025rfn}
\begin{equation}
    \Omega^{\mathrm{Spin}}_7 (B \mathbb{Z}_p) = \mathbb{Z}_p \oplus \mathbb{Z}_p \,, \quad ( p >3 \,, \text{prime}) \,.
\end{equation}
The generators for the two factors are given by a seven-dimensional lens space $L^7_p$ with $\mathbb{Z}_p$ gauge bundle and a three-dimensional lens space $L^3_p$ times K3 (here the gauge bundle is non-trivial only on the lens space), which can be thought of as capturing pure gauge and mixed gauge-gravitational anomalies. The anomaly theory can be written in terms of reduced Dirac $\eta$-invariants (where we subtract the pure gravitational part) and is given by
\begin{equation}
    \widetilde{\mathcal{A}}^F = \sum_q n_q \, \widetilde{\eta}^{\mathrm{D}}_q \,,
\end{equation}
with $n_q$ being the multiplicity of charge $q$ hypermultiplets. The theory is anomaly free if the anomaly theory evaluates to integers on both generators, i.e., 
\begin{equation}
    \widetilde{\mathcal{A}}^F [L^7_p] \in \mathbb{Z} \,, \quad \widetilde{\mathcal{A}}^F [L^3_p \times \mathrm{K3}] \in \mathbb{Z} \,,
    \label{eq:fermionanomfree}
\end{equation}
and anomalous otherwise. Given that $p$ is a prime bigger than $3$, these are both `mod $p$' conditions, which depend on $q^2$ and $q^4$ of the charged hypermultiplets. More concretely one has, \cite{Monnier:2018nfs, Hsieh:2020jpj, DOStoappear}
\begin{equation}
\begin{split}
    \widetilde{\mathcal{A}}^F [L^7_p] &= - \Big( \sum_{q = 1}^{p-1} n_q q^2 \Big) \Big( \frac{p^2-1}{24p} \Big) + \Big( \frac{1}{12} \sum_{q=1}^{p-1} n_q (q^2 - q^4) \Big) \Big( \frac{1-p}{2p} \Big) \,, \\
    \widetilde{\mathcal{A}}^F [L^3_p \times K3] &= - \Big( \sum_{q=1}^{p-1} n_q q^2 \Big) \Big( \frac{1-p}{p} \Big) \,.
\end{split}
\end{equation}
If there is an anomaly one can attempt to cancel it using a discrete version of the six-dimensional Green-Schwarz mechanism, where the chiral 2-forms also transform non-trivially under the discrete gauge symmetries. Via anomaly inflow this implies that the strings, coupling to the 2-form fields, host chiral fermions which are charged under the discrete symmetry. With this, the consistency of some of the models with discrete fermion anomalies can be restored. But the Green-Schwarz term can only cancel a subgroup of all fermion anomalies, as described in \cite{Monnier:2018nfs, Dierigl:2022zll, Dierigl:2025rfn, Cheng:2025ikd, DOStoappear}\footnote{See also \cite{Lee:2022spd, Saito:2025idl}.}. Instead of the evaluation of the fermionic anomaly theory to be integer as in \eqref{eq:fermionanomfree}, the new requirement is that it matches the Green-Schwarz contribution parameterized by $\tfrac{1}{p}R_i$ mod $\mathbb{Z}$:
\begin{equation}
    \widetilde{\mathcal{A}}^F [L^7_p] - \tfrac{1}{p} R_1 \in \mathbb{Z} \,, \quad \widetilde{\mathcal{A}}^F [L^3_p \times \mathrm{K3}] - \tfrac{1}{p} R_2 \in \mathbb{Z} \,,
\end{equation}
The value of $R_i \in \mathbb{Z}$ depends in the choice of quadratic refinement and the discrete anomaly inflow onto the string worldsheet (see \cite{Dierigl:2022zll, Dierigl:2025rfn, DOStoappear} for some concrete realizations). 

We now see the issue; satisfying the anomaly conditions requires a choice of fermion spectrum $\{n_q\}$ and will provide a lower bound on the number of charged hypermultiplets. Since the total number of hypermultiplets is finite, see \cite{Kim:2024eoa}, this can provide a bound on the allowed orders of discrete gauge groups $\mathbb{Z}_p$ if the number of charged hypermultiplets has to be very large. For prime $p$ the generic expectation is, that in order to satisfy the anomaly cancellation conditions one needs $\mathcal{O}(p)$ hypermultiplets.

Let us illustrate that for an example; $p =13$ with no Green-Schwarz contribution. Then one finds\footnote{Note that since $q$ enters quadratically, one gets the same result for $\pm q$ and we can focus on the range $q \in \{ 1 \,, \dots \,, \tfrac{p-1}{2} \}$.}
\begin{equation}
\begin{array}{c | c c c c c c}
q & 1 & 2 & 3 & 4 & 5 & 6 \\ \hline
q^2 \mod 13 & 1 & 4 & 9 & 3 & 12 & 10 \\
\tfrac{1}{12} (q^2 - q^4) \mod 13 & 0 & 12 & 7 & 6 & 2 & 12
\end{array}
\end{equation}
The anomaly constraints in this case simply reduce to 
\begin{equation}
    \sum_q n_q q^2 = 0 \mod 13 \,, \quad \frac{1}{12}\sum_q n_q (q^2 - q^4) = 0 \mod 13 \,.
\end{equation}
For only a single charge $q$ we see that we need at least $p$ charged hypermultiplets, but even combining different charges one obtains a lower bound. Here, one obtains an anomaly free spectrum with 3 hypermultiplets chosen as $n_1 = n_3 = n_4 =1$, but there is no anomaly free solution with two hypermultiplets, that satisfies both anomaly equations.

Similar arguments can be made in the case of general $\mathbb{Z}_N$ gauge groups with $N$ not necessarily prime, where the structure of the bordism groups and generators becomes more involved, \cite{Dierigl:2025rfn, DOStoappear}. However, while this argument can provide a lower bound for the number of charged hypermultiplets for fixed $N$ and Green-Schwarz contribution $R_i$, it is unfortunately not strong enough to provide a universal bound for $N$, which can be seen via the following counterexample.

We choose $N=p$ to be a prime of the form
\begin{equation}
    p = 8k +1 \,,
\end{equation}
of which there are infinitely many. In this case there is an integer $q$ such that $q^4 = -1$ mod $p$. Taking a hypermultiplet spectrum of the form
\begin{equation}
   n_r = n_{r^2} = n_{r^3} = n_{r^4} = 1\,,
\end{equation}
and all other $n_q$ vanishing, can be shown to solve the anomaly equation (in the absence of a Green-Schwarz term). Thus, these models only require four charged hypers but allow for arbitrary large orders of the gauge group $\mathbb{Z}_p$.

\section{Conclusions and future directions}
\label{sec:concl}

In this paper we have put a bound on the order of elements in discrete gauge symmetries for supersymmetric theories with 8 or more supercharges in various dimensions. With 8 supercharges we did not completely settle the discrete gauge symmetries where hypermultiplets carry the discrete charge, but instead put bounds on it assuming the symmetry acted by permutation. In many cases we were able to obtain sharp upper bounds which were realized by concrete string theory constructions.

One area of future directions would be to complete the story for hypermultiplets in theories with 8 supercharges. Another direction is to reduce the dimension. Indeed, if the conjecture in \cite{Kim:2024eoa} is correct the upper bound for the number of vectors comes from the toroidal compactification of the maximal tensor theory in 6d, in which case we can bound the order of discrete symmetries carried by vector multiplets in $D=4,5$ by using this. For example we would deduce that in $D=4$ the discrete element should belong to Sp$(492;\mathbb{Z})$ which leads to an upper bound. It would also be interesting to generalize these results to the cases with lower supersymmetry and in particular first to 4 supercharges in 4d, and then to the case without any supersymmetries. These cases would naturally be harder.  Nevertheless, the results of this paper reinforce the Swampland perspective on the boundedness of the order of discrete gauge symmetries in all cases.

Improved bounds on hypermultiplets with discrete charges would additionally be very interesting, since they may be naturally related to $\mathrm{U}(1)$ gauge theories with large charges via (un)Higgsing transitions. For that reason we expect that a strict upper bound on the discrete gauge symmetries of this type should shed light on the maximal $\mathrm{U}(1)$ charges and has to potential to rule out infinite families, such as the ones discussed in \cite{Taylor:2018khc}.

\section*{Acknowledgments}

The work of ZKB, CV and KX is supported in part by a grant from the Simons Foundation (602883, CV) and the DellaPietra Foundation. HK is supported by the National
Research Foundation of Korea (NRF) grant funded by the Korean government (MSIT) (2023R1A2C1006542).
The work of MD is funded by the European Union (ERC, SymQuaG, 101163591). MD and HK further thank the Harvard Swampland Initiative and its Visitors Program during which this work has been initiated.

\begin{appendix}

\section{Maximal order from compact subgroups}
\label{app:compact}

In this appendix we collect some details of the approach to derive bounds on the maximal order and maximal prime order element in the exceptional U-duality groups E$_{k(k)}(\mathbb{Z})$, from elements in the simply-connected cover of the maximal compact subgroup $K^D_{\mathrm{U}}$. In all of the interesting cases the action of the maximal compact subgroup within the U-duality group over $\mathbb{R}$ contains the antisymmetric representation, which we will use to extract bounds on the maximal order elements.

Given an element in the defining representation of the simply connected cover of $K^{D}_{\mathrm{U}}$, with eigenvalues $\lambda_i$, the same element in the antisymmetric representation has eigenvalues $\lambda_i \lambda_j$ $(i<j)$. Using the discussion of maximal order elements associated to the companion matrices of cyclotomic polynomials in Section \ref{subsec:rough}, one has to make sure that the resulting element in $E_{k(k)}(\mathbb{R})$ contains all necessary eigenvalues. In particular, if $\lambda_i \lambda_j = e^{2 \pi i \frac{a}{n_{ij}}}$, the resulting elements needs to contain all eigenvalues of the form $e^{2 \pi i \frac{r}{n_{ij}}}$, with $r$ coprime to $n_{ij}$ at a fixed multiplicity. This therefore fixes at least $\phi(n_{ij})$ eigenvalues of the element in $\mathrm{E}_{k(k)}(\mathbb{R})$. For $D \in \{ 3 \,, 4 \}$ the branching in \eqref{eq:compactrep} allows a combination of eigenvalues with those of other representations, while we focus on bounds from the antisymmetric representation only.

Since it is not automatic that the element is conjugate to an integer element in $G^D_{\mathrm{U}}$ the bounds derived in this way will in general not be sharp. For the maximal prime order in $D \in \{3,4\}$, however, the bounds are guaranteed to be realized.

\subsection{Maximal prime order}

For the maximal prime order $p$ we can use the general approach described in \cite{steinberg1968lectures, Serre_Robinson_MacLachlan_Wall_1979, 2010arXiv1011.0346S} to derive bounds. These can be obtained by considering the U-duality groups not over $\mathbb{Z}$ but over the finite field $\mathbb{F}_s$. The order of the associated group is given in terms of the number of positive roots, $A$, and the basic degrees, $d_i$, of the Weyl-invariant polynomials. For $\mathrm{E}_{k(k)}(\mathbb{F}_s)$ it is given by, e.g., \cite{steinberg1968lectures},
\begin{equation}
	|\mathrm{E}_{k(k)}(\mathbb{F}_s)| = s^A \prod_i (s^{d_i} - 1) \,.
\end{equation}
Then one considers the map
\begin{equation}
	\mathrm{E}_{k(k)}(\mathbb{Z}) \rightarrow \mathrm{E}_{k(k)}(\mathbb{F}_s) \,.
\end{equation}
Given some integral element $g$ with prime order $p$, this gives a nontrivial order $p$ element modulo $s$ for all but finitely many prime numbers $s$. This implies that $p$ divides $ s^A \prod_i (s^{d_i} - 1)$. If $d_i<p-1$ for all $i$, we may first pick $a$ so that $a^{d_i}\neq 1 \mod{p}$ and then pick a prime number $s$ so that $s=kp+a$ (there are infinitely many primes $s$ of this form), and for this $s$, $p$ does not divide $ s^A \prod_i (s^{d_i} - 1)$. So we see that $d_{max}\geq p-1$ for the maximal prime order. Actually, we have $d_{max}= p-1$ because the orbifold Euler characteristic of the moduli space has $p$ in the denominator, which implies the existence of an orbifold point with unbroken discrete gauge symmetry with order $p$, see, e.g., \cite{2010arXiv1011.0346S}, for $D \in \{ 3,4\}$. In the case of $\mathrm{E}_{6(6)}(\mathbb{Z})$ the Euler characteristic vanishes, but the evidence below suggests the existence of a $\mathbb{Z}_{13}$ subgroup.

As an example let us discuss the case of E$_{8(8)}(\mathbb{Z})$ for which one has
\begin{equation}
    A = 120 \,, \enspace d_i \in \{ 2\,, 8 \,, 12 \,, 14 \,, 18 \,, 20 \,, 24 \,, 30\} \,,
\end{equation}
and setting $s = 2$ one finds that $|\mathrm{E}_{8(8)} (\mathbb{F}_2)|$ contains a factor $(2^{30}-1)$ which is divisible by 31 and already suggests the existence of a $\mathbb{Z}_{31}$. That such a subgroup indeed exists can then be shown by determining the Euler characteristic, which for $\mathrm{E}_{8(8)}(\mathbb{Z})$ has denominator given by
\begin{equation}
    2^{30} \cdot 3^{13} \cdot 5^5 \cdot 7^4 \cdot 11^2 \cdot 13^2 \cdot 19 \cdot 31 \,,  
\end{equation}
and for example guarantees the existence of a $\mathbb{Z}_{31}$ as well as a $\mathbb{Z}_{19}$ subgroup.\footnote{It is interesting to note that there is no $\mathbb{Z}_{17}$ subgroup, the reason is that the eigenvalues of the antisymmetric and spinor representation cannot combine in such a way to produce all $e^{2 \pi i \frac{r}{17}}$ at fixed multiplicity.}

For E$_{k(k)}(\mathbb{Z})$ with $k \in \{ 6, 7\}$, we can even be more explicit by providing the eigenvalues of the element in the defining representation of the simply connected cover of $K^D_{\mathrm{U}}$. Denoting
\begin{equation}
    \xi_{p} = e^{2 \pi i/p} \,,
\end{equation}
one can choose
\begin{equation}
    \lambda_i \in \{ \xi_{13} \,, \xi_{13}^{12} \,, \xi_{13}^{2} \,, \xi_{13}^{11} \,, \xi_{13}^{3} \,, \xi_{13}^{10} \,, \xi_{13}^5 \,, \xi_{13}^{8}\}
\end{equation}
for the four pairs of complex conjugate eigenvalues in USp$(8)$. This produces all $\xi_{13}^r$ at fixed multiplicity 2 in the antisymmetric representation embedded in E$_{6(6)}(\mathbb{R})$. For $k = 7$ we can choose the 8 eigenvalues of SU$(8)$ to be given by
\begin{equation}
    \lambda_i \in \{ \xi_{19}^7 \,, \xi_{19}^9 \,, \xi_{19}^{10} \,, \xi_{19}^{11} \,, \xi_{19}^{13} \,, \xi_{19}^{14} \,, \xi_{19}^{15} \,, \xi_{19}^{16} \} \,.
\end{equation}
combining $\{ \lambda_i \lambda_j: i<j\}$ with its complex conjugates, then results in $\xi_{19}^r$ at fixed multiplicity 3, where here it is curical that the branching contains the antisymmetric as well as its complex conjugate.

\subsection{Maximal order}

To bound the maximal order elements we start with an element of the simply-connected cover of the maximal compact subgroup $K^{D}_{\mathrm{U}}$, which has complex eigenvalues $\lambda_i$ given by roots of unity
\begin{equation}
	\lambda_i = e^{2 \pi i \frac{a_i}{n_i}} \,,
\end{equation}
where we assume the fractions $\frac{a_i}{n_i}$ to be fully reduced. The antisymmetric representation then has the set of eigenvalues
\begin{equation}
	\big\{ \lambda_i \lambda_j = \text{exp} \big( 2 \pi i \tfrac{a_i n_j + n_j a_i}{n_i n_j}\big) : \enspace i < j \big\} \,.
    \label{eq:antisymeigenv}
\end{equation}
After reducing the fraction we collect the set denominators in the antisymmetric representation $\{ n_{ij}\}$. The order of the resulting element is given by
\begin{equation}
    N = \text{lcm} \{ n_{ij} \} \,,
\end{equation}
The requirement that the resulting matrix can be conjugated to a block-diagonal if cyclotomic companion matrices then requires the appearance of all eigenvalues of the form $e^{2 \pi i \tfrac{r}{n_{ij}}}$, with $r$ coprime to $n_{ij}$ and thus implies a lower bound on the size of the matrix. Since one can embed $\mathrm{E}_{k(k)}(\mathbb{Z})$ into GL$(\text{dim}(\mathbf{R});\mathbb{Z})$ consistency requires
\begin{equation}
    \sum_{\{ n_{ij}\}} \phi(n_{ij}) \leq \text{dim}(\mathbf{R}) \,.
    \label{eq:antiboundgen}
\end{equation}
The bounds can then be obtained by maximizing $N$ while staying within the bound \eqref{eq:antiboundgen}. Note that since in $D \in \{3,4\}$ the branching involves also other representations these bounds are rather conservative and not necessarily saturated by physical models.

The derivation of such bounds is made significantly harder due to the fact that there can be cancellations between the numerator and denominator in \eqref{eq:antisymeigenv}. These can only appear in case the $n_i$ are not coprime. In these cases some of the summands in \eqref{eq:antiboundgen} can become smaller while generically others become larger. In order to make the search algorithm efficient we therefore restrict to eigenvalues $\lambda_i$ such that the set of denominators $\{ n_i\}$, where we do not count the same denominator twice, is a set of coprime integers. The bounds are then derived under the following assumption, which we did not find a counterexample for.

\vspace{0.2cm}

{\bf Coprime assumption:} We assume that the maximal order $N$ in the antisymmetric representation can be achieved by choosing a coprime decomposition $\{n_i\}$ referring to the set of denominators appearing for the eigenvalues in the defining representation of the simply-connected version of the maximal compact subgroup.

\vspace{0.2cm}

Under this assumption there are various simplifications. In particular one has $n_{ij} = n_i n_j$, implying 
\begin{equation}
    N = \text{lcm} \{ n_i \} \,,
\end{equation}
and one can use the multiplicity properties of the Euler totient to simplify
\begin{equation}
    \sum_{i < j} \phi (n_{ij}) = \sum_{i < j} \phi(n_i) \phi(n_j) \,.
\end{equation}
As mentioned above, while examples suggest that the assumption is satisfied we do not have a general proof that this is always the case.

For E$_{6(6)}(\mathbb{Z})$ we note that we start with four pairs of complex conjugate eigenvalues, which demands that the antisymmetric representation only contains 24 non-trivial eigenvalues and one has
\begin{equation}
    \sum_{i<j} \phi(n_i n_j) = \sum_{i<j} \phi(n_i) \phi(n_j) \leq 24 \,,
\end{equation}
where we used the multiplicative property for $\phi$ for coprime numbers. The bound of the maximal order that can be achieved in this way is $N = 90$, with the sum given by exactly $24$. In this case $\phi(90) = 24$ and it would be consistent to take $\lambda_i = \xi_{90}^r$. However, since 90 is even all $r$ need to be odd, which is not possible in the antisymmetric, thus in this case we can exclude 90 and all other even $N$ with $\phi(N) = 24$. This leads to the refined bound
\begin{equation}
    N \leq 60 \,,
\end{equation}
which is actually realized. A specific choice of eigenvalues $\lambda_i$ is given by
\begin{equation}
    \{ \lambda_i \} = \{ \xi_{12} \,, \xi_{12}^5 \,, \xi_{12}^7 \,, \xi_{12}^{11} \,, \xi_5 \,, \xi_5^2 \,, \xi_5^3 \,, \xi_5^4 \} \,, 
\end{equation}
ordering the antisymmetric products $\lambda_i \lambda_j$ we find the eigenvalues of the element in the antisymmetric representation
\begin{equation}
\begin{split}
    \{ \xi_{12}^6 (\mathbb{Z}_2); \xi_{12}^4, \xi_{12}^8 (\mathbb{Z}_3) ; \xi_{12}^6 (\mathbb{Z}_2); \xi_5, \xi_5^2, \xi_5^3, \xi_5^4 (\mathbb{Z}_5); \xi_{12}^r \xi_5^s (\mathbb{Z}_{60}); 1, 1, 1, 1 \} \,,
\end{split}
\end{equation}
which is of order $60$.

In the case of E$_{7(7)}(\mathbb{Z})$ the decomposition also involves the complex conjugate of the antisymmetric representation and thus we do not expect our bounds to be realized. The same strategy as above leads to 
\begin{equation}
    N \leq 240 \,,
\end{equation}
with the prime decomposition $\{16 \,, 3 \,, 5\}$ leading to
\begin{equation}
    \sum_{i<j} \phi(n_i) \phi(n_j) = \phi(16) \phi(3) + \phi(16) \phi(5) + \phi(3) \phi(5) = 16 + 24 + 8 = 48 \leq \text{dim}(\mathbf{R}) = 56 \,.
\end{equation}
The first value we know can be realized is given by $N = 210$, which might very well be the sharp upper bound.

For E$_{8(8)}(\mathbb{Z})$ the decomposition further involves the spinor representation, which will not enter our bounds at all. The upper bound we derived is given by
\begin{equation}
    N \leq 6930 \,,
\end{equation}
with the decomposition
\begin{equation}
    \{ n_i \} = \{ 5 \,, 7 \,, 9 \,, 22\} \,,
\end{equation}
leading to 
\begin{equation}
    \sum_{i<j} \phi(n_i n_j) = 244 \,. 
\end{equation}
As mentioned above, since we do not use the spinor representation we do not expect this bound to be realized and the largest realized is given by $N = 420$.

\section{6d anomaly cancellation}\label{app:6danomaly}

In six-dimensional supergravity, chiral fields generate one-loop gauge and gravitational anomalies. These anomalies can be canceled through the Green–Schwarz–Sagnotti mechanism \cite{Green:1984bx,Green:1984sg,Sagnotti:1992qw}. The anomaly cancellation then requires factorization of the one-loop anomaly polynomial as
\begin{align}\label{eq:anomaly-polynomial}
    I_8 = \frac{1}{2} \,\Omega_{\alpha\beta} X^\alpha X^\beta \ , \quad 
    X^\alpha = -\frac{1}{2} b_0^\alpha \,\mathrm{tr} R^2 + \frac{1}{2}\sum_i \,\frac{b_i^\alpha}{\lambda_i} \mathrm{tr} F_i^2 \ ,
\end{align}
where $R$ is the curvature two-form, $F_i$ is the field strength of a non-Abelian factor $G_i$, and $\lambda_i$ denotes the group-theory normalization constant summarized in Table \ref{tb:G-normalization}. The matrix $\Omega$ denotes the tensor intersection form with signature $(1,T)$, and $b_0, b_i$ are anomaly vectors in $\mathbb{R}^{1,T}$, which also need to satisfy certain quantization conditions, see, e.g., \cite{Monnier:2017oqd, Cheng:2025ikd}.

This factorization is possible only when the massless spectrum and the anomaly vectors $b_0, b_i$, and $\Omega$ satisfy the following conditions:
\begin{align}\label{eq:GS-conds-non-abelian}
&H - V = 273 - 29 T,\qquad b_0^2 = 9 - T, \qquad
b_0\cdot b_i = \frac{\lambda_i}{6}\left(\sum_{\bf R} n_{\bf R}^i A_{\bf R}^i - A^i_{\bf adj}\right), \\
&B_{\bf adj}^i = \sum_{\bf R} n^i_{\bf R} B^i_{\bf R} \,,\enspace
b_i\cdot b_i = \frac{\lambda_i^2}{3}\left(\sum_{\bf R} n^i_{\bf R} C^i_{\bf R} - C^i_{\bf adj}\right), \enspace b_i\cdot b_j = 2 \lambda_i \lambda_j \sum_{\bf R,S} n_{\bf R,S}^{ij} A^i_{\bf R} A^j_{\bf S}, \enspace i\neq j, \nonumber
\end{align}
where $n^i_{\bf R}$ ($n^{ij}_{\bf R,S}$) is the multiplicity of matter in representation ${\bf R}$ ($({\bf R, S})$) of the gauge group $G_i$ ($G_i \times G_j$), and the trace indices $A^i_{\bf R}, B^i_{\bf R}, C^i_{\bf R}$ are defined by
\begin{align}
    {\rm tr}_{\bf R}F^2=A_{\bf R}{\rm tr}F^2 , \quad {\rm tr}_{\bf R}F^4 = B_{\bf R}{\rm tr}F^4 + C_{\bf R}({\rm tr}F^2)^2 \ .
\end{align}
The first condition in \eq{eq:GS-conds-non-abelian} is the gravitational anomaly cancellation condition, which we extensively use in the discussion of Section \ref{sec:8}.

\begin{table}[t]
\centering
\begin{tabular}{|c|c|c|c|c|c|c|c|c|}
    \hline
    $G_i$ & $\mathrm{SU}(N)$ & $\mathrm{SO}(N)$ & $\mathrm{Sp}(N)$ & $\mathrm{G}_2$ & $\mathrm{F}_4$ & $\mathrm{E}_6$ & $\mathrm{E}_7$ & $\mathrm{E}_8$ \\
    \hline 
    $\lambda_i$ & $1$ & $2$ & $1$ & 2 & 6 & 6 & 12 & 60 \\ 
    \hline
\end{tabular}
\caption{Group-theory normalization constants for non-Abelian gauge algebras.}
\label{tb:G-normalization}
\end{table}

\end{appendix}

\bibliography{papers}{}
\bibliographystyle{JHEP} 

\end{document}